\documentclass[]{aa1}		 

\usepackage{natbib}
\bibpunct{(}{)}{;}{a}{}{,}
\usepackage{lscape}
\usepackage{graphicx}
\usepackage{longtable}
\usepackage{supertabular}
\usepackage{txfonts}
\usepackage{color}

\definecolor{Red}{rgb}{1, 0, 0}
\definecolor{Green}{rgb}{0, 1, 0}
\definecolor{Blue}{rgb}{0, 0, 1}
\definecolor{Black}{rgb}{0, 0, 0}
\definecolor{White}{rgb}{1, 1, 1}
\definecolor{Grey}{rgb}{0.5, 0.5, 0.5}
\definecolor{Yellow}{rgb}{1, 1, 0}
\definecolor{Magenta}{rgb}{1, 0, 1}
\definecolor{Cyan}{rgb}{0, 1, 1}
\definecolor{Orange}{rgb}{0.8, 0.3, 0}
\definecolor{DarkGreen}{rgb}{0.2, 0.7, 0.1}
\definecolor{Pink}{rgb}{1, 0.4, 0.7}

\begin{document}

\title{Atomic and molecular gas from the epoch of \\ reionization down to redshift 2}

\titlerunning{Cold gas and {\sc ColdSIM} }

\authorrunning{AA.VV.}

\author{Umberto Maio,\inst{1,2}  C{\'e}line P{\'e}roux,\inst{3,4}  Benedetta Ciardi\inst{2}}

\offprints{U. Maio, e-mail: umberto.maio@inaf.it}

\institute{
    INAF-Italian National Institute of Astrophysics, Observatory of Trieste, via G. Tiepolo 11, 34143 Trieste, Italy
	\and
	Max Planck Institute for Astrophysics, Karl-Schwarzschild-Str. 1, 85748 Garching bei M\"unchen, Germany
	\and
    European Southern Observatory, Karl-Schwarzschild-Str. 2, 85748 Garching bei M\"unchen, Germany; 
	\and
    Aix Marseille Universit\'e, CNRS, LAM (Laboratoire d'Astrophysique de Marseille) UMR 7326, 13388, Marseille, France
}


\abstract
{Cosmic gas makes up about 90\% of baryonic matter in the Universe and H$_2$ is the closest molecule to star formation.}
{In this work we study cold neutral gas and its H$_2$ component at different epochs, exploiting state-of-the-art hydrodynamic simulations that include time-dependent atomic and molecular non-equilibrium chemistry coupled to star formation, feedback effects, different UV backgrounds presented in the recent literature and a number of additional processes occurring during structure formation ({\sc ColdSIM}).
}
{We predict gas evolution and contrast the mass density parameters and gas depletion timescales, as well as their relation to cosmic expansion, in light of the latest IR and (sub-)mm observations in the redshift range $ 2 \lesssim z \lesssim 7$.
}
{By performing updated non-equilibrium chemistry calculations we are able to broadly reproduce the latest HI and H$_2$ observations.
We find neutral-gas mass density parameters $ \Omega_{\rm neutral} \simeq  10^{-3}$ and increasing from lower to higher redshift, in agreement with available HI data.
Because of the typically low metallicities during the epoch of reionization, time-dependent H$_2$ formation is mainly led by the H$^-$ channel in self-shielded gas, while H$_2$ grain catalysis becomes important in locally enriched sites at any redshift.
UV radiation provides free electrons and facilitates H$_2$ build-up while heating cold metal-poor environments.
Resulting H$_2$ fractions can be as high as $\sim 50\%$ of the cold gas mass at $z\sim 4$-8, in line with the latest measurements from high-redshift galaxies.
The H$_2$ mass density parameter increases with time until a plateau of $ \Omega_{\rm H_2} \simeq 10^{-4}$ is reached.
Quantitatively, we find an agreement between the derived $ \Omega_{\rm H_2} $ values and the observations up to $z\sim 7$ and both HI and H$_2$ trends are better reproduced by our non-equilibrium H$_2$-based star formation modelling.
The predicted gas depletion timescales decrease at lower $z$ in the whole time interval considered, with H$_2$ depletion times remaining below the Hubble time and comparable to the dynamical time at all $z$.
This implies that non-equilibrium molecular cooling is efficient at driving cold-gas collapse in a broad variety of environments and since the very early cosmic epochs.
While the evolution of chemical species is clearly affected by the details of the UV background and gas self-shielding, the assumptions on the adopted initial mass function, different parameterizations of H$_2$ dust grain catalysis, photoelectric heating and cosmic-ray heating can affect the results in a non-trivial way.
In appendix, we show detailed analyses of individual processes, as well as simple numerical parameterizations and fits to account for them.
}
{Our findings suggest that, in addition to HI, non-equilibrium H$_2$ observations are pivotal probes for assessing cold-gas cosmic abundances and the role of UV background radiation at different epochs.}

\maketitle



\section{Introduction}\label{Sect:introduction}


\noindent
About 90\% of the baryons in the Universe are in the form of gas and most of this resides either in the cosmic space, constituting the so-called inter-galactic medium (IGM), or around galaxies, where it shapes the circum-galactic medium (CGM).
Only a tiny, but extremely relevant, fraction of gas populates neutral high-density regions, where the local medium cools and fragments. There, a large portion of the chemical composition is represented by neutral hydrogen, HI, and molecular hydrogen, H$_2$, the most abundant molecule in the Universe.
Different gas phases are not necessarily separated, as structure formation brings diffuse material into dense molecular regions, as well as spreads initially dense parcels into the hotter CGM/IGM.
How the involved processes work in detail is still matter of debate and the interplay between different phases has drastic consequences for the ionization state of H and the global amount of H$_2$.
\\
In the whole $z < 6$ epoch, the measured neutral-gas mass density normalized to the present-day critical density (neutral-gas density parameter) is $\Omega_{\rm neutral} \sim 10^{-3}$ and increases towards higher redshifts
\cite[][]{Noterdaeme2009, Zafar2013, Rhee2013, Sanchez2016, Jones2018, Hu2019, Walz2021, Chen2021}.
H$_2$ molecules have a more complex behaviour and are notoriously more difficult to probe.
Until a couple of years ago there was little reliable information at intermediate redshifts and no data at all about cold-gas H$_2$ content in the distant Universe.
Only recently first data from a number of observational campaigns based on e.g.
VLA,\footnote{\url{http://www.vla.nrao.edu/}}
ALMA\footnote{\url{https://www.almaobservatory.org}},
NOEMA\footnote{\url{https://www.iram-institute.org/EN/noema-project.php}} or
UKIRT\footnote{http://wsa.roe.ac.uk/UKIRT.html
}
have started to set constraints on its cosmological evolution.
\\
The exact processes driving molecular build-up in different cosmological environments are not fully understood, yet.
For this reason, H$_2$ mass in targeted objects is usually inferred through empirical fits, based, for instance, on emission from 
CO \cite[][]{Scoville1975, Dickman1986, Maloney1988, Hughes2017}
or 
[C II] \cite[][]{Madden1997, Pak1998, HT1999}
calibrated in the local Universe.
Heavy elements and C-rich molecules have been detected in metal-enriched collapsing sites, suggesting that CO cores exist surrounded by a [C II] envelope where H$_2$ is self-shielded from photodissociating radiation \cite[][]{Roellig2006, Wolfire2010, Genzel2012, Bolatto2013, Zanella2018, Madden2020, Gong2020}.
Notwithstanding the uncertainties related to calibration techniques, completeness and conversion factors, the currently estimated H$_2$ molecular-mass density in the local Universe is $ \rho_{\rm H_2} \simeq 6.8 \times 10^6 \, \rm M_\odot/Mpc^3 $  and accounts for about 20 per cent of the total abundance of cold gas \cite[][]{Fletcher2021}.
Despite the large scatter of one order of magnitude or more among different studies, the H$_2$ mass density parameter 
$ \Omega_{\rm H_2} $
is observed to increase at $ z\simeq 0$-3, assembling its mass with practically no observed environmental dependence 
\cite[][]{Darvish2018, Tadaki2019, Garratt2021},
and to mimick the cosmological star formation rate (SFR) density at higher $z$
\cite[][]{Decarli2020, Riechers2020vlaspec}.
These observations have stressed the longstanding need to explore the origin of the large supply of H$_2$ gas required to sustain star formation at different $z$
\cite[][]{Rodighiero2019, Tacconi2020, Hunt2020}
and questioned the ability to explain the large H$_2$ fractions (above 50\%) detected in galaxies around $z\sim 4$-6 \cite[][]{DZ2020, Boogaard2021}, in a primordial, presumably metal-poor, gas.\\
From a theoretical point of view, cold gas plays a key role for cooling and fragmentation.
It hosts the conversion from atomic to molecular phase, through which it fuels collapse and star formation
(\citealt{SaZi67}, \citealt{Peebles_Dicke68}).
Despite the well-determined HI observational trend, a full understanding of its physical behaviour over cosmological history still needs to be consolidated.
H-derived molecules are crucial for gas chemistry evolution \cite[see e.g.][for reviews]{BY2011, Girichidis2020}, because they supply low-temperature coolants to cosmic gas and are the main drivers of gas depletion timescales.
A full assessment of H$_2$ evolution in different epochs is elusive, because of both the lack of observational data at our disposal until recently and the large computing power needed to perform detailed chemistry calculations within three-dimensional simulations.
A correct evaluation of cold-gas fractions is sensitive to a number of cooling and heating processes due to either small-scale physics or large-scale events \cite[][]{Maio2021}.
Numerical analyses typically predict lower-than-observed neutral/HI mass densities and have difficulties in reproducing their 
increasing trend in redshift at $z\simeq 2$-6.
Accurate chemistry calculations following H$_2$ formation and the consequent run-away cooling regime require a full time-dependent "non-equilibrium" approach that is computationally expensive and hence not widespread within numerical implementations of cosmological simulations.
It has so far been employed to study mainly primordial epochs \cite[][]{Abel1997, Yoshida2003, Maio2010, Wise2012, Muratov2013, Skinner2020}, while non-equilibrium evaluations for $\Omega_{\rm neutral}$, $\Omega_{\rm H_2}$ and depletion times during and after the epoch of reionization are still missing.\\
The main channels for H$_2$ production in pristine media, composed mainly by $\rm H$, He and electrons, are based on catalysis of intermediate species interacting with H, with possible photon emission accompanying recombination \cite[][]{GalliPalla1998},
i.e. 
H$^-$ catalysis 
($\rm  H + e^- \rightarrow H^-  + \gamma $ and $\rm H^-  + H \rightarrow H_2 + e^- $)
and 
H$_2^+$ catalysis 
($\rm H + H^+ \rightarrow H_2^+ + \gamma $ and $ \rm H_2^+ + H  \rightarrow H_2 + H^+ $).
While these processes are dominant at intermediate densities below $10^4\,\rm K$ or in recombining gas where free charges are available, three-body interactions ($ \rm 3H \rightarrow H_2+H$) could further contribute in dense neutral media.
\\
For metal-enriched environments the picture is even more complex, since metallicity ($Z$) evolution and dust grain catalysis might influence H$_2$ abundances
($\rm 2H \rightarrow H_2 $ on dust grains), 
also depending on the efficiency of photoelectric heating
\cite[e.g.][]{BakesTielens1994}, 
whose implications on H$_2$ production at different cosmological epochs are currently uncertain.
Cosmic-ray protons losing their energy in dense molecular gas
\cite[][]{GoldsmithLanger1978}
or in thinner photodissociating regions 
\cite[][]{Shaw2009}
could heat the medium and affect early H$_2$ formation in a non-trivial way.
Their possible role for pristine primordial gas has been pointed out with basic semi-analytic calculations by \cite{Jasche2007}.
However, also because of the poor information on cosmic-ray heating yield and rate at that time, their impact has not been simulated within three-dimensional numerical simulations of early structure formation and the implications on (non-equilibrium) HI and H$_2$ masses are still unknown.
\\
Feedback mechanisms can remove local gas mass and inject entropy into the surrounding medium \cite[e.g.][]{Nagamine2004}.
As UV photons are capable to dissociate molecules and ionize neutral atoms in non-shielded regions, the build-up of a UV background determined by forming structures in the first Gyr, at redshift $z \gtrsim 6$, affects the gas thermal state, as well as HI and H$_2$ density parameters.
Gas self-shielding
\cite[e.g.][]{DB1996, Rahmati2013, Hartwig2015, WG2017, Ploeckinger2020, Luo2020, Walter2021},
limits the ability of UV radiation to penetrate dense material and preserves HI or H$_2$ fractions in star forming regions.
The relevance and the relative role of the two main (HI and H$_2$) shielding processes for chemical abundances in cosmic environments need to be further explored, though.
\\
In addition, stellar evolution modulates the whole baryon cycle depending on stellar initial mass function (IMF), metal yields and lifetimes.
The shape of the IMF has implications for the resulting stellar population, its chemical patterns and the galaxy luminosity function
\cite[][]{Leitherer1999, BC2003, Vazdekis2010, Tescari2014, Mancini2015, Mancini2016, Ma2015, Ma2017, Valentini2019}.
\\
Modifications in the supernova (SN) efficiency and/or in the mechanical feedback can affect the whole energetics of the picture at all times and induce variations in the amount of gas mass expelled from the collapsing regions into the IGM
\cite[][]{Maio2011, Campisi2011},
with possible consequences on atomic and molecular species (in particular $\Omega_{\rm neutral}$ and $\Omega_{\rm H_2}$).
\\
More or less powerful sources alter the ionization state of the IGM and suppress gas cooling depending on the strength of radiative feedback and HI or H$_2$ shielding parameterisations
\cite[][]{DB1996, Oshea2008, PS2011, Petkova2012, Wise2012prad, Rahmati2013, Rahmati2015, Sternberg2014, Maio2016, WG2017, JB2018}.
\\
How much all these processes affect the overall molecular chemistry in realistic environments and at different redshifts is matter of debate 
\cite[][]{Genel2014, Furlong2015, Pallottini2017, Dave2017, Dave2019, Donnari2019}.
In this respect, dedicated numerical efforts to explore the role of the various physical phenomena that condition the manifold regimes of cosmic environments must be performed to interpret available data.
Throughout this work we will shed light on the evolution of cosmic atomic and molecular gas by interpreting new, recent, state-of-the-art observations and by modelling HI and H$_2$ species in the frame of high-resolution N-body hydrodynamical simulations that include full time-dependent non-equilibrium chemistry, star formation feedback and stellar evolution.
A further goal of our study is to show how neutral and molecular gas densities and corresponding depletion times are modulated by different UV radiation fields, HI or H$_2$ gas self-shielding, stellar population parameters and the local processes affecting cosmic chemistry in different epochs.
\\
We adopt a flat $\rm\Lambda$CDM cosmological model, consistently with cosmic microwave background data \cite[][]{Komatsu2011}, with present-day expansion parameter normalised to 100~$\rm km/s/Mpc$ of 
$h = 0.7$.    
Baryon, matter and cosmological-constant density parameters are assumed to be
$\Omega_{0,\rm{b}} = 0.045$, 
$\Omega_{0,\rm{m} } = 0.274$    
and
$\Omega_{0,\rm{\Lambda}} = 0.726$,     
respectively.
The spectral parameters are
$\sigma_8 = 0.8 $    
for the $z=0$ mass variance within 8~$\rm Mpc/{\it h}$ radius and 
$n=0.968$ 
for the slope of the primordial power spectrum, consistent with the latest Planck results
\cite[][]{Planck2020}.
The present-day cosmological critical density is 
$\rho_{0, \rm{crit} } \simeq 277.4 \, h^2\, \rm M_\odot/kpc^3$.
\\
The paper is organized as follows.
Details on observational data, numerical implementations and data analysis are given in Sect.~\ref{Sect:method};
results are presented in Sect.~\ref{Sect:results} and discussed in Sect.~\ref{Sect:discussion}; conclusions are summarised in Sect.~\ref{Sect:conclusions}.


\section{Method}\label{Sect:method}


In the following we present the most recent observations of cold atomic and molecular gas, the theoretical models implemented (comprising several physical processes newly included in the simulations), as well as a comprehensive list of all our runs.

\subsection{Observations}
\label{sect:data}

To quantify the amount of HI and H$_2$ masses at various epochs, different authors use either mass densities, $\rho_{\rm HI}$ and $\rho_{\rm H_2}$ (in $\rm M_\odot/Mpc^3$), or the corresponding dimensionless $\Omega_{\rm HI}$ and $\Omega_{\rm H_2}$ mass density parameters.
The evolution of the entire amount of cold (below $10^4\,\rm K$) neutral gas density, $ \rho_{\rm neutral}$, can be obtained, as usually done in the literature, correcting H-based measurements by helium content, i.e. scaling HI masses by a factor of 1.3 \cite[e.g.][]{Crighton2015}.
In the following we will discuss neutral and molecular gas in terms of
$\Omega_{\rm neutral} = \rho_{\rm neutral} / \rho_{0,\rm{crit} } $ and
$\Omega_{\rm H_2} = \rho_{\rm H_2} / \rho_{0, \rm{crit} } $.
In practice, these parameters correspond to the mass fraction of the considered species times $\Omega_{0,\rm{b} }$. \\
We consider HI data at $z \lesssim 5$, as collected by \cite{PerouxHowk2020} (PH20) in their Tab.~1, which rely on 
low-$z$ 21~cm emission ($ z \lesssim 0.4 $) and high-$z$ quasar absorption data.
At low redshift, 21~cm emission analysis and spectral stacking are performed on data from 
2dFGRS \cite[][]{Delhaize2013},
WSRT \cite[][]{Rhee2013, Hu2019},
GMRT \cite[][]{Rhee2016, Rhee2018, Bera2019},
Arecibo/ALFA \cite[][]{Hoppman2015} and 
ALFALFA \cite[][]{Jones2018}
observational programs.
The present-day estimated value of the HI mass density parameter is
$ \simeq (4.02 \pm 0.26) \times 10^{-4} $ 
\cite[][]{Hu2019}.
High-redshift measurements are obtained using data from high-resolution spectroscopic surveys of tens of objects performed at 
VLT \cite[][]{Zafar2013},
Calar Alto/CALIFA \cite[][]{Sanchez2016},
HST \cite[][]{Rao2017},
and lower-resolution spectra of hundreds or thousands of objects recorded by SDSS \cite[][]{Noterdaeme2009, Noterdaeme2012, Crighton2015}. 
The HI density evolution is quite smooth, increasing by a factor of $\sim 3 $ from the present up to $z\simeq 5.3$
(see Fig.~\ref{fig:OmegaUVdata}).
Mass density parameters for cold neutral gas vary within 
$\sim 0.5$ and $\sim 1.4 \times 10^{-3}$ 
and can be fitted by eq.~(13) in 
\cite{PerouxHowk2020}, i.e.: 
$ \Omega_{\rm neutral} = ( 4.6 \pm 0.2) \times 10^{-4} \, (1+z)^{0.57 \pm 0.04}  $.
\\
Up-to-date H$_2$ mass densities at different $z$
(for any given observational programme) are found in
\cite{Riechers2020vlaspec, Riechers2020coldz} (VLASPEC, COLDz),
\cite{Decarli2020} (ASPECS),
\cite{Lenkic2020} (PHIBBS-2),
\cite{Garratt2021} (UKIDSS-UDS),
and 
 \cite{Hamanowicz2021}
(ALMACAL-CO).
Some of these are also present in the compilation by
\cite{PerouxHowk2020}.
Upper limits at $z\simeq 0$-2 are provided by
\cite{Klitsch2019} (ALMACAL-absorption, ALMACAL-abs hereafter) and impose
$ \rho_{\rm H_2} < 10^{8.39} \,\rm M_\odot/Mpc^3 $,
corresponding to $\Omega_{\rm H_2} < 1.81 \times 10^{-3} $.
This value is much higher than early Herschel PEP lower limits of molecular-mass density parameters $ \gtrsim 10^{-5}$ 
inferred through indirect methods for gas at $z\lesssim 3$ \cite[][]{Berta2013}.
The currently established present-day value of 
$\rho_{\rm H_2} \simeq 6.8 \times 10^6 \, \rm M_\odot/Mpc^3$ 
is obtained by \cite{Fletcher2021} (xCOLDGASS) and corresponds to 
$ \Omega_{\rm H_2} \simeq 5 \times 10^{-5}$
with a factor-of-2 relative error.
ALMA and VLA observational programmes have determined an increasing 
$ \Omega_{\rm H_2} $ trend from such $z \simeq 0$ value to values $\sim 10^{-4}$ at $ z\simeq 2$-3
\cite[][]{Decarli2020, Riechers2020vlaspec}.
Infrared UKIDSS-UDS constraints
\cite[][]{Garratt2021}
have confirmed existing data, finding molecular-mass densities 
$\rho_{\rm H_2} \simeq 2 \times 10^{7} \rm M_\odot/Mpc^3$ (i.e. $\Omega_{\rm H_2} \simeq 1.5 \times 10^{-4}$) at $z\simeq 2$.
H$_2$ fractions up to 50-60\% in galaxies at redshift $z\sim 4$-6 have been lately reported by e.g. \cite{DZ2020}, although
H$_2$ densities within 1$\sigma$ confidence levels decrease down to 
$ \Omega_{\rm H_2} \sim 10^{-5} $ towards $z \simeq 7$
\cite[][]{Riechers2020coldz}.
The derived $\Omega_{\rm H_2}$ values at low $z$ span more than two orders of magnitude, from $ 10^{-6}$ to few $10^{-4}$, but they are more robust at $z \gtrsim 2$.
In practice, the overall ensemble of recent $\Omega_{\rm H_2} $ observational findings show a variety of results comprised between 
$ 10^{-3} $ and 
$ 10^{-6}$ in the 
$z \simeq 0$-7 range and the shape of the resulting $\Omega_{\rm H_2}$ evolution is completely different from the smooth $\Omega_{\rm neutral}$ one.
A list with the detailed values
derived by different observational programs is presented in Appendix  \ref{AppendixData} (Tab.~\ref{tab:dataOmegaH2}).
\\
From $ \rho_{\rm neutral } $ and $ \rho_{\rm H_2} $, it is possible to derive the corresponding depletion times,
$t_{\rm depl, neutral} = \rho_{\rm neutral} /  \dot\rho_{\star} $
and
$t_{\rm depl, H_2} = \rho_{\rm H_2} /  \dot\rho_{\star} $, where
$\dot\rho_{\star}$ is the cosmic star formation rate density.

\subsection{Theoretical models}

To perform a solid interpretation of observational data, we run and analyse a number of numerical simulations, based on N-body and hydrodynamical calculations with an extended version of the parallel code {\sc P-Gadget3} \cite[][]{Springel2005} modified to address in detail cold-gas atomic and molecular content over different cosmological epochs ({\sc ColdSIM}).
Besides gravity and smoothed-particle hydrodynamics, the code contains a self-consistent time-dependent treatment of atomic and molecular chemistry for e$^-$, H, H$^+$, H$^-$, He, He$^+$, He$^{++}$, H$_2$, H$^+_2$, D, D$^+$, HD, HeH$^+$, which follows, in non-equilibrium conditions, a network of reactions for species ionization, recombination and dissociation
\cite[][]{Abel1997, Yoshida2003, Maio2007}.
Here the network has been extended to account for the main processes taking place from the epoch of reionization down to lower redshift and these are coupled to star formation, stellar evolution and heavy-element enrichment of He, C, N, O, Ne, Mg, Si, S, Ca, Fe, etc. from SN~II, AGB and SN~Ia phases 
\cite[][]{Tornatore2007, Tornatore2010, Maio2010, Maio2011}.
In the following sections we describe in more detail the chemical processes included in the original version of the code, as well as the additional processes included for a better modelling of cold-gas evolution.

\subsubsection{Basic chemistry evolution}

Non-equilibrium chemistry is followed through a set of first-order differential equations accounting for creation and destruction processes, as well as corresponding heating or cooling, as described in e.g. \cite{Maio2010} and \cite{Maio2016}.
Number density evolution for each species and for each gas particle is computed by solving the differential equations through a backward difference scheme with an integration time of 0.1 the actual timestep \cite[to grant numerical convergence;][]{Anninos1997}.\\
Stellar evolution is followed for a range of stellar masses and initial metallicities.
Stars with masses above $8\,\rm M_\odot$ explode as SNe~II and inject a typical kinetic energy of $10^{51}\,\rm erg$ in the surrounding medium, while lower-mass stars evolve through AGB or SNe~Ia phase with consequent mass loss \cite[][]{Tornatore2007}.
Star formation is modelled stochastically \cite[e.g.][]{Katz1992, SpringelHernquist2003} in particles with density above a threshold of $10 \, \rm cm^{-3}$ and temperature below $10^4\,\rm K$ after the "loitering" regime and the catastrophic run-away cooling phase \cite[][]{Maio2009}.
Wind feedback takes place with a reference wind velocity of 350~km/s, while wind particles are decoupled from hydrodynamics and treated in a collisionless fashion \cite[i.e. they do not cool;][]{Maio2011} for a delay time after the initial kick of 0.025-0.1 the Hubble time, $t_{\rm H}$.
Spreading of heavy elements during different stellar phases is traced for He, C, N, O, Ne, Mg, Si, S, Ca, Fe, etc. Metal-dependent yields are taken from \cite{WW1995}, \cite{vandenHoek1997}, \cite{Woosley2002} and \cite{Thielemann2003}, while stellar lifetimes follow \cite{Padovani1993}
\cite[see also Sect.~2.1 in][for more details]{Maio2016}.
The reference initial mass function (IMF) has a \cite{Salpeter1955} shape over the stellar-mass range ${\rm [0.1, 100]~M_\odot}$.
\\
The chemical network is coupled to the insurgence of UV radiation according to available $z$-dependent tabulated photoheating and photoionization rates.
These are added to the local gas heating caused by ongoing exothermic chemical reactions and physical processes.
Works in the recent literature have suggested different epochs for the establishment of a UV background and different temporal evolution of the adopted UV rates, which we test by means of dedicated runs.
The main physical differences among the UV models are related to the treatment of the possible sources of cosmic re-ionization and the availability of observational data to constrain model parameters at various $z$.
We adopt the standard literature prescriptions by \cite{HM1996} (HM), as well as the most up-to-date modelling including latest high-$z$ data by \cite{P2019} (P19) and \cite{FG2020} (FG20).
While the HM case is a simple model based on early QSO data only, the FG20 model include also the contributions from early galaxies with an onset of HI and HeII photoheating at $z\simeq 8$ and 4, respectively (similarly to the \citealt{HM2012}, where HI and HeII photoheating kicks in at $z\simeq 9$ and 5) and seems to reproduce better the Ly$\alpha$ forest mean flux decrement in the low-redshift IGM \cite[][]{Christiansen2020}.
HM rates have a fairly smooth evolution until $z\simeq 6$, although updated analyses by P19, including contributions from both QSOs and galaxies, suggest rates that have a tail extending up to $z\simeq 15$.
 \\
Although this set-up is already enough to follow the time-dependent HI and H$_2$ evolution with a precision that is not commonly obtained in standard implementations of cosmic structure formation models, 
to evaluate even more accurately the properties of cold gas in different environments, additional processes that might influence chemical abundances in complex ways should be included, such as:
HI and H$_2$ self-shielding from UV radiation,
H$_2$ grain catalysis,
photoelectric as well as 
cosmic-ray heating.
In the following we will discuss separately all these newly implemented processes.

\subsubsection{Gas HI and H$_2$ self-shielding}

\noindent
While UV radiation can easily penetrate thin media, gas self-shielding effects should be taken into account when estimating HI and H$_2$ fractions in dense gas.
\\
For H$_2$ self-shielding we adopt the formulae by \cite{DB1996}, who fit their results either by means of a single power law (their eq.~36) or by a slightly more accurate expression (their eq.~37).
We also use the more recent results of the work by \cite{WG2017}, which, similarly to \cite{Sternberg2014}, is a reformulation of the ones by \cite{DB1996}.
Most runs discussed here include H$_2$ shielding according to eq.~36 of \cite{DB1996} if not otherwise stated.
In the case of \cite{WG2017}, the optimal scale to estimate the local H$_2$ column density, $ N_{\rm H_2} $, and related shielding have been investigated in depth by \cite{Luo2020}.
They showed that $ N_{\rm H_2} $ can be quantified as
$ N_{\rm H_2} = n_{\rm H_2} \lambda_{\rm J } / 4 $,
where 
$ n_{\rm H_2} $ is the the three-dimensional H$_2$ number density and 
$\lambda_{\rm J}$ the Jeans length.
This has been proven to be an excellent approximation.
In Appendix \ref{AppendixShielding} we show that different H$_2$ shielding implementations have a negligible effect on our results.
\\
For HI self-shielding we rely on the redshift and density dependent corrections proposed by \cite{Rahmati2013}.
We note that, while the authors provide corrections at $z \le 5$, both observational data and theoretical investigations suggest the presence of cold shielded gas also at higher redshifts.
Despite the numerical uncertainties discussed in the original paper, the redshift evolution of self-shielding corrections as a function of density presents only little (non-monotonic) variations and deviations from the $z$-averaged mean/median values for any given density are modest.
Therefore, to capture the essential behaviour of HI, we fit the median values of those shielding curves.
As the redshift dependence is weak and in the absence of a better prescription, we extend such relation to $ z > 5 $.
The resulting density-dependent fitting expression is given in Appendix \ref{AppendixShielding} (eq.~\ref{fshield_fit}).
This nicely shows that gas with H number densities $n_{\rm H} \lesssim 10^{-3}\,\rm cm^{-3}$ is essentially un-shielded (i.e. the correction factor is close to unity), while at larger number densities the effective photoionization rate decreases by several orders of magnitude.
\\
Thus, while we find that the implementation of gas self-shielding in the simulations affects the resulting $\Omega_{\rm neutral}$ and $\Omega_{\rm H_2}$ (this will be discussed in detail in Sect.~\ref{Sect:results}), variations in its exact formulation have a minor impact.

\subsubsection{H$_2$ dust grain catalysis}

\noindent
We have coupled H$_2$ gas-grain catalysis to the non-equilibrium chemical network and followed it for different gas temperatures and metallicities.
Condensation of metals into dust grains can enhance H$_2$ formation via $\rm H $ abstraction, although the quantum modelling of H$_2$ formation on dust grains bears uncertainties of a factor of 2 \cite[][]{Bron2014}.
Chemical rates for H$_2$ gas-grain catalysis are expected to be of the order of $\sim 10^{-17}\,\rm cm^3/s $
\cite[][]{Duley1996}
and generally scale with the sticking coefficient (i.e. the fraction of particles that adsorb to the grain surface), $S_{\rm H}$, and the square root of gas temperature, $\sqrt{T}$.
\cite{Cazaux2004} suggested a rate of $3 \times 10^{-17} \, \rm cm^3/s$ and a sticking coefficient following a second-order polynomial depending on the adopted grain temperature, $ T_{\rm gr} $ 
\cite[][]{HM1979}.
Based on \cite{TH1985}, \cite{Omukai2005} propose an H$_2$ rate of $ 6 \times 10^{-17} \,\rm cm^3/s$ for gas at 300~K (corresponding to about $3.5 \times 10^{-17} \, \rm cm^3/s$ at 100~K) and the same sticking coefficient as \cite{HM1979}.
Similar approaches have been used later on e.g. by \cite{Tomassetti2015} and more recently by \cite{Sillero2021}.
Updated quantum-mechanical calculations by
\cite{Thi2020}
find a value of $3.74 \times 10^{-17} \,\rm cm^3/s$ for a gas at 80~K assuming a constant $ S_{\rm H} $ of unity.
Broadly speaking, all these results are in line with observational constraints of 3-$4.5 \times 10^{-17} \,\rm cm^3/s $
\cite[][]{Jura1975a, Jura1975b, Gry2002},
although the value by \cite{Thi2020} does not incorporate a temperature-dependent $ S_{\rm H} $.
Gas-grain energy losses scale proportionally to $ \sqrt{T} \, (T-T_{\rm gr}) $, hence contributing to gas heating or cooling depending on both $T$ and $T_{\rm gr} $.
On the base of Milky Way calibrations, the normalization is expected to be $\sim 10^{-33} \,\rm erg/s/K^{3/2} cm^3 $, with a roughly 1-dex error at low $Z$.
Such large deviations introduce a remarkable degree of freedom in the parameter study, suggesting for example that variations of the dust-to-gas ratio within a factor of 10 would be still consistent with the constraints.
For this reason, the exact shape and $Z$ scaling of dust-induced H$_2$ formation and cooling/heating rates have been long debated in the literature
\cite[][]{HM1989, Omukai2005, Draine2007, Goldsmith2011, Krumholz2011, GloverClark2012}.
\\
Here, we adopt an H$_2$ formation rate of $3.5 \times 10^{-17} \, \rm cm^3/s $ at 100~K, scaled by $ \sqrt{T} $ and $S_{\rm H}$ 
\cite[][]{HM1979, Cazaux2004} 
(see Appendix \ref{AppendixH2gr}) to update the H$_2$ abundances employed in the cooling/heating calculations \cite[][]{Omukai2005}.
Since observational determinations are based mainly on data from the Milky Way, we scale the rates linearly with $Z$ at metallicities different from solar.
As a fiducial value we assume $T_{\rm gr} = 75\,\rm K$ at all $z$.
Nevertheless, we additionally check alternative hypotheses (see Appendix \ref{AppendixH2gr}), i.e. a lower and a higher value of 40~K and 120~K, respectively, both consistent with observational constraints and theoretical estimates of grain properties\footnote{
Local determinations of $ T_{\rm gr} $ are related to gas surrounding nearby early-type stars (60-120~K) and photodissociation regions (65~K for $\theta^1$ Ori~C), instead theoretical estimates refer to H astration modeling on a perfect surface (75~K) or completely imperfect surface (44~K).
Updated observational best-fit values at different redshifts are
$\sim 41$~K at $z\simeq 4.5$ and 38-43~K at $z\simeq 5.5$
\cite[][]{Khusanova2020, Faisst2020}.
}.
To study the effects of an evolving $T_{\rm gr}$, we also consider either grain temperatures varying as the CMB temperature, i.e. 
$ T_{\rm gr} (z) = T_{\rm 0,CMB} (1+z) $ with 
$ T_{\rm 0,CMB} \simeq 2.7\,\rm K $,
or varying in accordance to the energy balance between CMB radiation and dust grain power-law emission \cite[][]{DraineLee1984}, i.e.
$ 
T_{\rm gr}^{4+\beta}(z) = 
    T_{\rm 0,gr}^{4+\beta} +  
    T_{\rm 0,CMB}^{4+\beta} [ (1+z)^{4+\beta} -1 ],
$
with 
$\beta \simeq 2$ emissivity index\footnote{
 Common $ \beta $ values are between roughly 1 and slightly greater than 2, but the resulting $T_{\rm gr} (z)$ is not very sensitive to particular choices.}
chosen in accordance to the latest ALMA determinations \cite[][]{daCunha2021} and 
$T_{\rm 0,gr} = 18\,\rm K$ 
present grain temperature \cite[][]{daCunha2013}.
We explore degeneracies and metallicity effects at different cosmological epochs implementing also the possibility to adopt fixed values for $Z$
(e.g. 0.01-$1 \,\rm Z_\odot$).\\
Different $T_{\rm gr}$, either constant or $z$-dependent, cause local fluctuations of comparable amplitude on H$_2$ chemistry, but have little impact on the global $\Omega_{\rm H_2}$ behaviour.
Metallicity is relevant for values close to $Z_\odot$, which boost both local H$_2$ fractions (by one dex since $z\simeq 8$) and density parameters (by more than two dex at $z\gtrsim 10$ and by a factor of a few at $z\simeq 4$).
\\
In practice, H$_2$ grain catalysis has a relevant local impact in metal enriched sites, while, generally speaking, its global role decreases at higher $z$ as a result of the decreasing metallicity.
Finally, it does not affect the global HI chemical history and $\Omega_{\rm neutral}$ values do not vary significantly from the reference case.

\subsubsection{Photoelectric heating}

Gas-grain interactions are usually accompanied by photoelectric heating.
This is one of the main heating processes of the diffuse interstellar medium (ISM) and is caused by the absorption of UV photons by dust grains which then re-emit electrons and heat up the surrounding gas.
According to standard expressions commonly used in ISM investigations
\cite[][]{Weingartner2001}, the photoelectric heating rate density can be written as a function of 
gas temperature, 
Habing parameter $G_0$ quantifying the local UV field \citep{Habing1968} 
and electron fraction
\cite[][]{DraineSutin1987, BakesTielens1994}.
With the exception of Milky Way calibrations, the details of such process in different cosmological contexts are still unknown.
We thus either consider the fiducial constant case of $G_0 = 1.7$, or, because of the close relation between star formation and UV photon production, we scale it by the local SFR
\cite[][]{NarayanKrumholz2017}, 
i.e. 
$ G_0 \rightarrow G_0 \times \rm SFR/SFR_{\odot}$, 
where
$\rm SFR_{\odot} \sim 2 \, M_\odot/yr $ is the value of the Galactic SFR \cite[][]{Chomiuk2011, Licquia2015}.\footnote{
A precise quantification of the local SFR depends on the IMF and different assumptions can lead to errors up to $40\%$. This is smaller than typical uncertainties of some chemistry rates (known within up to a factor of 2), hence induced variations should not be dramatic. 
}
We have verified that the HI and H$_2$ densities obtained with both prescriptions are very similar, with differences visible only on local scales (see details in Appendix~\ref{AppendixPE}).
We note that, while heating the surrounding gas, the photoelectric effect provides additional free electrons, useful for the H$^-$ channel.
These charges react with H via radiative attachment and increase H$_2$ formation via associative detachment. \\
The H$_2$ abundances reached in the presence of photoelectric heating are higher by a factor of a few mostly at early times, while the overall effect on $ \Omega_{\rm neutral} $ is modest.
\\

\subsubsection{Cosmic-ray heating}

\noindent
Another source that likely plays a role for H$_2$ evolution is the cosmic-ray heating.
Cosmic rays are mostly constituted by high-energy protons usually formed in explosive SN events that can impact on molecular-mass build-up during and after the epoch of reionization.
It is known that heating by cosmic rays is related to ionization of atoms and consequent production of free electrons. Each cosmic-ray proton ionizes the medium generating slow electrons and these lead to second or later-generation ionization in 2/3 of the cases (causing an increase in the yield of liberated energy by 5/3)
\cite[][]{Spitzer1969, Opal1971}.
About $50\%$ of the energy released goes into gas heating.
Because of their relatively large abundances, the species mainly involved in the ionization process and accounting for most of the gas heating are H, He and H$_2$
\cite[][]{Glassgold2012},
while others are almost irrelevant\footnote{
The probability that cosmic-ray protons interact with e.g. dust or heavy molecules is expected to be orders of magnitude smaller than interactions with H$_2$ molecules. As a reference, the dust cross section per H nucleus in the diffuse ISM is $\sim 10^{-21}\,\rm cm^{-2}$, while in the case of H$_2$ it is $\sim 10^{-17}\,\rm cm^{2}$. Even if interactions with dust should occur, only $1\%$ of them would produce photoelectrons 
\cite[][]{DwekSmith1996}.
}.
Gas heating due to cosmic rays is usually parameterized as a function of the heating yield and ionization rate.
A wide range of values exists in the literature for the cosmic-ray heating yield, which is estimated to be $\sim 7$-20~eV
\cite[][respectively]{StahlerPalla2004, GoldsmithLanger1978}
when considering electron energies higher than 1~keV, but reach about 44~eV if including also lower electron energies
\cite[][]{Glassgold2012}.
As the normalisation of the transferred energy is weakly dependent on the local environment, we always assume 20~eV as typical value.
The ionization rate is expected to be 
$ \zeta \sim 10^{-17} \, \rm s^{-1}$ per H atom, although IR measurements of 
$\zeta \gtrsim 10^{-16} {\rm s}^{-1} $ 
in dense molecular sites
\cite[][]{Indriolo2007, IndrioloMcCall2012, Hollenbach2012}
demonstrated that it can vary with density and reach larger values.
New analysis by \cite{Thi2020} has suggested a typical ionization rate of 
$1.7 \times 10^{-17} \, \rm s^{-1}$ per H$_2$ molecule.
Recent theoretical results by \cite{Padovani2018} have shown that $\zeta$ could be described by a decreasing function of the density, peaking at a value of $\sim 10^{-16}\,\rm s^{-1}$ at 
gas column densities of $10^{19}\, \rm cm^{-2}$.
This estimate is more than a factor of ten larger than the one by
\cite{Thi2020}, but still consistent with SN energies of $10^{51} \,\rm erg$ and with uncertainties related to the different derivation technique.
\\
In our implementation we couple cosmic-ray heating with non-equilibrium chemistry to handle both a constant $\zeta = 10^{-17} \,\rm s^{-1}$ and a density-dependent $ \zeta $, like the one expected by \cite{Padovani2018}.
In this latter case, we fit tabulated results of their model $\mathcal{L}$ in a friendly form 
(given in Appendix \ref{AppendixZeta}).
To constrain extreme behaviours, cosmic-ray heating is applied either to all gas particles in the simulated volume or to star forming particles only.
Because of the lack of data for ionization rates in regimes different from the local one, $\zeta$ is generally scaled by the local SFR, as it is linked to SN shocks in star forming regions.
\\
While we refer the reader to Appendix \ref{AppendixZeta} for more details on the impact of different prescriptions, here we just mention that the inclusion of our reference cosmic-ray heating (i.e. a density-dependent $\zeta$ applied to star forming particles) affects only mildly the density parameters at all redshifts.

\subsubsection{Star formation model}
\label{SFmodel}

Finally, commonly used star formation implementations that convert gas into stars stochastically estimate the SFR relying on the ratio between the gas density above a given threshold, $\rho_{\rm th}$, and a typical star formation time
that scales with gas density. 
In particular, the star formation timescale can be written as 
$ t_{\rm sf} = t_{\rm sf, 0}  ( \rho/ \rho_{\rm th} )^{1/2} $, 
where the maximum star formation time $  t_{\rm sf, 0} $ is an empirical parameter that varies substantially in different observations.
At each integration time, collisionless star particles are spawned stochastically with a probability given by 
$ p =  (m/m_{\rm star}) \{ 1 - \exp[ - (1 - \beta) x \Delta t / t_{\rm sf}  ] \} $,
with 
$m$ gas particle, 
$m_{\rm star}$ stellar particle, 
$ \beta $ fraction of short-lived stars that die as SNae,
$ x $ cold fraction and 
$ \Delta t $ numerical timestep 
\cite[][]{SpringelHernquist2003}. 
The resulting star formation rate associated to each gas parcel is then
$ \dot M_{\rm sf} = (1 - \beta) x m / t_{\rm sf} $.
Here we adopt a reference value of $  t_{\rm sf, 0}  = 2~ \rm Gyr $ 
(in line with \citealt{Leroy2013, Tacconi2013, Tacconi2020}).
The value of 
$ m_{\rm star} $ is estimated by assuming that each gas particle forms 4 stellar particles (4 `generations'),
$ \beta $ is computed according to the adopted IMF and 
$ x $ follows \cite{SpringelHernquist2003}, resulting usually close to unity in high-density regions.
In practice, all this means that at the threshold value the actual star formation time is $ t_{\rm sf} = t_{\rm sf, 0} $ and the star formation probability $p$ is small, being $ \Delta t / t_{\rm sf} \ll 1$ typically.
At larger densities, $ t_{\rm sf} $ becomes shorter, while the star formation probability $p$ increases.
Since we are able to follow detailed non-equilibrium H$_2$ evolution, we additionally implement an H$_2$-based approach in which the SFR is driven by the non-equilibrium H$_2$ density in each gas parcel above the star formation threshold of $10~\rm cm^{-3}$.
In general, the H$_2$ star formation efficiency can be thought as implicitly coded in the star formation timescale, thus efficiency values equal to (lower than) unity correspond to timescale values equal to (higher than) the reference timescale. 
We run corresponding simulations and discuss the implications for HI and H$_2$ evolution in the next sections.

\subsection{Numerical simulations}

\noindent
All the processes mentioned above and newly implemented in our theoretical framework have an impact on the amount of free electrons or protons available for H$_2$ formation, in addition to increasing gas heating or cooling and associated species destruction or formation.
What their net result is in different environments and at different epochs needs to be investigated.
\\
To this goal, we run a number of numerical simulations in a cubic volume of length 10 co-moving Mpc/{\it h} 
({\sc ColdSIM}), as listed in Tab.~\ref{tab:properties}.
Gas and dark-matter fields are each sampled with 512$^3$ particles at an initial redshift of 100
\cite[different initial redshifts have a modest effect on the results; e.g.][]{Maio2011vb}.
This set-up corresponds to a resolution of
$7.89 \times 10^4\,\rm M_\odot$ and 
$4.72 \times 10^5\,\rm M_\odot$,
for gas and dark-matter particles, respectively.
In our reference simulation (HM-HISSmed in Tab.~\ref{tab:properties}) star formation happens in gas above 10~cm$^{-3}$ 
with a \cite{Salpeter1955} IMF.
It has a SN efficiency of 0.1, 
wind velocity of 350~km/s,
wind delay time of $ 0.025 \, t_{\rm H}$,
HM UV background, 
density-dependent HI self-shielding according to 
eq. \ref{fshield_fit} 
and H$_2$ self-shielding by \cite{DB1996} (their eq. 36).
Additional simulations have been run to explore and quantify the role of different assumptions (for UV background, star formation and feedback effects related to IMF, SNe or winds), as well as gas physics and chemistry processes (HI or H$_2$ shielding, H$_2$ formation channels, H$_2$ dust grain processes, cosmic-ray heating), as detailed in Tab.~\ref{tab:properties}.
Implications from the different HM, P19 and FG20 backgrounds (highlighted with different colors in the table and in the next figures) have been addressed with dedicated HI-shielded or un-shielded simulations.
We note that most HI-shielded runs include HI self-shielding according to eq.~\ref{fshield_fit} (and are denoted by the HISSmed suffix after the UV background name, i.e. 
HM-HISSmed, P19-HISSmed and FG20-HISSmed), 
while HI self-shielding according to \cite{Rahmati2013}'s tables is tested only in selected runs (denoted via the HISS suffix after the UV background name, i.e.
HM-HISS, P19-HISS and FG20-HISS).
The gas physics and chemistry modelling described above has been investigated by always using the HM background.
For example, this choice has been done when adopting (non-equilibrium) H$_2$-based star formation (HM-HISSmed-H2).
Although our fiducial star formation threshold is $10\,\rm cm^{-3}$, comparisons have been made to investigate the impact on the H$_2$ mass density parameter of a standard low-density threshold $\sim 0.1\,\rm cm^{-3}$ with an HM background (HM-std and HM-HISSmed-std), finding that in this latter case H$_2$ fractions are not resolved properly.
Similarly, we have adopted the HM scenario to quantify the contribution of each H$_2$ formation channel by including in the reference run the three-body processes together with the H$^-$ and H$_2^+$ channels (HM-HISSmed-3b).
H$_2$ self-shielding by \cite{DB1996} (eq.~37) or \cite{WG2017} are tested via HM-HISSmed-3b-DB2 and HM-HISSmed-3b-WG runs.
To isolate the implications of individual channels, two further tests have been implemented, the first including the H$_2^+$ channel only (HM-HISSmed-H$_2^+$) and the second including none of the H$_2$ channels (HM-HISSmed-none).
A similar analysis has been done for H$_2$ grain catalysis at different grain temperatures and metallicities (see Tab.~\ref{tab:properties}), 
with or without HI shielding
(HM-HISSmed-cat-75-Zevol, HM-cat-75-Zevol), 
or photoelectric heating
(HM-HISSmed-cat-75-pe).
Cosmic-ray heating with constant or density-dependent \cite[][]{Padovani2018} ionization rate in shielded gas has been explored by means of HM-HISSmed-cr and HM-HISSmed-crP18 simulations, respectively. To have a more realistic analysis, we additionally consider the possibility that cosmic-ray heating is injected only in star forming parcels (HM-HISSmed-crP18sfr).
Effects linked to variations of the SN efficiency and IMF have been tested by simply employing un-shielded UV backgrounds to maximize the impact of stellar feedback.
Other features related to powerful winds (at 700~km/s) or winds re-coupled after a longer time (wind delay time of $0.1\,t_{\rm H}$, 4 times the reference $0.025\,t_{\rm H}$) are explored, too.
For each output time, cosmic structures are identified by means of a friends-of-friends algorithm with a linking length of $20$ per cent the mean inter-particle separation. Substructures are identified by the Subfind algorithm \cite[][and references therein]{Dolag2009} and are post-processed to trace: masses, positions, radii, velocities, star formation rates, 
temperatures, chemical abundances (of e$^-$, H, H$^+$, H$^-$, He, He$^+$, He$^{++}$, H$_2$, H$^+_2$, D, D$^+$, HD, HeH$^+$, C, N, O, Ne, Mg, Si, S, Ca, Fe, etc.), substructures and all the relevant physical properties of each object.
Our choices for initial conditions, box size and resolution are determined by the necessary trade-off between the required accuracy of the physical descriptions implemented in the code and the numerical feasibility of the runs.

\begin{table*}  
\small
\centering
\caption{
Inputs to simulation models.
}
\label{tab:properties}
\begin{tabular}{c | l | c c c c c c c}
\hline
\hline
 \textcolor{Grey}{Properties} & Simulation run	& UV &	HI shielding	&   H$_2$ shielding & IMF & $\varepsilon_{\rm SN}$ & Extra modules\\
 \textcolor{Grey}{\tiny{(a)}}   & 	\tiny{(b)}	&	\tiny{(c)}	&	\tiny{(d)}	&	\tiny{(e)}	&	\tiny{(f)}	&	\tiny{(g)}	&	\tiny{(h)} \\
\hline
 & HM						&	HM&	--					&	DB96	&	S	&	0.1	&	--\\ 
 & HM-HISS				&	HM&	R13		&	DB96	&	S	&	0.1	&	--\\
 & {\bf HM-HISSmed} (ref)		&	HM&	\ref{fshield_fit}&	DB96&	S	&	0.1	&	--\\
 & \textcolor{Blue}{P19}       &	P19	&	--					&	DB96	&	S	&	0.1	&	--\\
\textcolor{Grey}{UV rate}   & \textcolor{Blue}{P19-HISS}  &	P19	&	R13	&	DB96	&	S	&	0.1	&	-- \\
\textcolor{Grey}{modeling}  & \textcolor{Blue}{{\bf P19-HISSmed}}	&	P19	&	\ref{fshield_fit}   &	DB96&	S	&	0.1	&	-- \\
 & \textcolor{Orange}{FG20}		&	FG20 	&	--	&	DB96	&	S	&	0.1	& -- \\
 & \textcolor{Orange}{FG20-HISS}   &	FG20 	&	R13	&	DB96	&	S	&	0.1	&	--\\
 & \textcolor{Orange}{\bf FG20-HISSmed}&	FG20 	& \ref{fshield_fit}&	DB96	&	S	&	0.1	&	--\\
\hline
 & HM-std        &HM &	--  &	DB96	&	S	&	0.1	& threshold $ 0.1 \,\rm cm^{-3}$ \\
\textcolor{Grey}{star formation} & HM-HISSmed-std                          &HM & \ref{fshield_fit}&	DB96&	S	&	0.1	& threshold $ 0.1 \,\rm cm^{-3}$ \\
 & \textcolor{Red}{\bf HM-HISSmed-H2}  &HM & \ref{fshield_fit}&	DB96&	S	&	0.1	& non-equilibrium H$_2$-based\\
\hline
 & HM-HISSmed-3b		&	HM&	 \ref{fshield_fit}&	DB96&	S	&	0.1	& 3-body processes \\
 & HM-HISSmed-3b-DB2	    &	HM& \ref{fshield_fit}&	DB96 (eq. 37) &	S	&	0.1	& 3-body processes \\
 & HM-HISSmed-3b-WG		&	HM&	 \ref{fshield_fit}&	WG17&	S	&	0.1	& 3-body processes \\
 & HM-HISSmed-H$_2^+$	&	HM&	 \ref{fshield_fit}&	DB96&	S	&	0.1	& H$_2^+$ channel only\\
 & HM-HISSmed-none		&	HM&	 \ref{fshield_fit}&	DB96&	S	&	0.1	& no H$_2$ channels\\
 & HM-HISSmed-cat-75-Zevol		&	HM&	 \ref{fshield_fit}&	DB96 &	S	&	0.1	& catalysis at 75~K \\
 & HM-HISSmed-cat-75-pe	&	HM&	 \ref{fshield_fit}&	DB96 &	S	&	0.1	& catalysis at 75~K and \\
  & & & & & & & photoelectric heating \\
\textcolor{Grey}{gas physics}
 & HM-HISS-cat-40-Zevol		&	HM&	R13 &	DB96&	S	&	0.1	& catalysis at 40~K \\
\textcolor{Grey}{\& chemistry}
 & HM-HISS-cat-75-Zevol		&	HM&	R13 &	DB96&	S	&	0.1	& catalysis at 75~K \\
 & HM-HISS-cat-120-Zevol		&	HM&	R13 &	DB96&	S	&	0.1	& catalysis at 120~K \\
 & HM-HISS-cat-Tcmb-Zevol	&	HM&	R13 &	DB96&	S	&	0.1	& catalysis at $T_{\rm CMB} (z)$ \\
 & HM-HISS-cat-Tbeta-Zevol	&	HM&	R13 &	DB96&	S	&	0.1	& catalysis at $T_\beta (z)$ \\
 & HM-HISS-cat-75-Zsun		&	HM&	R13 &	DB96&	S	&	0.1	& catalysis at 75~K and $Z = \rm Z_\odot$ \\
 & HM-HISS-cat-75-0.01Zsun	&	HM&	R13 &	DB96&	S	&	0.1	& catalysis at 75~K and $Z = 10^{-2} \rm Z_\odot $ \\
 & HM-HISS-cat-75-Zt	&	HM&	R13 &	DB96&	S	&	0.1	& catalysis at 75~K and $Z = Z(t) $ \\ 
 & HM-cat-75-Zevol			&	HM      &	-- 	&	DB96 &	S	&	0.1	& catalysis at 75~K \\
 & HM-HISSmed-cr		&	HM&	 \ref{fshield_fit}&	DB96&	S	&	0.1	& constant $\zeta$ , all gas\\
 & HM-HISSmed-cr-P18	&	HM&	 \ref{fshield_fit}&	DB96&	S	&	0.1	& $\zeta$ by P18,$^\ast$ all gas\\ 
 & HM-HISSmed-cr-P18sfr&	HM&	 \ref{fshield_fit}&	DB96&	S	&	0.1	& $\zeta$ by P18,$^\ast$  SFR-scaled\\\hline
 & HM-1			&	HM&	--				    &	DB96	&	S	&	1.0	& --\\
 & HM-0.5			&	HM&	--					&	DB96	&	S	&	0.5	& --\\
 & HM-0.01			&	HM&	--					&	DB96	&	S	&	0.01	& --\\
 & HM-w700 &	HM&	--	&	DB96	&	S	&	0.1	& wind velocity of 700~km/s\\ 
\textcolor{Grey}{feedback} &
HM-wf &	HM&	--			&	DB96	&	S	&	0.1	& winds decoupled for $0.1 \, t_{\rm H}$\\
\textcolor{Grey}{\& IMF} & 
 HM-Chabrier	&	HM&	--					&	DB96	&	Ch	&	0.1	& --\\
 & HM-1-Chabrier &	HM&	--					&	DB96	&	Ch	&	1.0	& --\\
 & \textcolor{Blue}{P19-1}	        &	P19	&   --  &	DB96	&	S	&	1.0	& --\\
 & \textcolor{Blue}{P19-Chabrier}	&	P19	&	--	&	DB96	&	Ch	&	0.1	& --\\
 & \textcolor{Blue}{P19-1-Chabrier}&	P19	&	--	&	DB96	&	Ch	&	1.0	& --\\
\hline
\hline
\end{tabular}
\begin{flushleft}
Blocks in the first column (a) collect, from top to bottom, runs with different
implementation properties.
Following columns refer to:
simulation name (b),
cosmological UV rates (c), 
HI shielding (d), 
H$_2$ shielding (e), 
IMF (f), 
SN energy fraction (g),
and 
description of extra modules adopted besides basics ones (h).
UV backgrounds are by 
\citet[][HM]{HM1996}, \citet[][P19]{P2019} or \citet[][FG20]{FG2020};
HI shielding by 
\citet[][R13]{Rahmati2013} or eq.~\ref{fshield_fit} in Appendix \ref{AppendixShielding};
H$_2$ shielding by 
\citet[][DB96, their eq. 36 if not otherwise specified]{DB1996} or \citet[][WG17]{WG2017};
IMF by 
\citet[][S]{Salpeter1955} or 
\citet[][Ch]{Chabrier2003}.
\\
$\ast$~Simple fit in Appendix \ref{AppendixZeta} to the $\mathcal{L}$ model by \cite{Padovani2018}.
\end{flushleft}
\end{table*}


\section{Results}\label{Sect:results}


In the following we present the main results obtained by the analysis of the simulations and their behaviour with respect to the observations described before.

\subsection{Gas thermal state}
\begin{figure*}
  \small{
    \hspace{2.7cm} \textcolor{Blue}{P19}
    \hspace{3.8cm} \textcolor{Orange}{FG20}
    \hspace{3.8cm} HM 
    }\\
  \vspace{-0.9cm}\\
  \centering
  \includegraphics[width=0.35\textwidth]{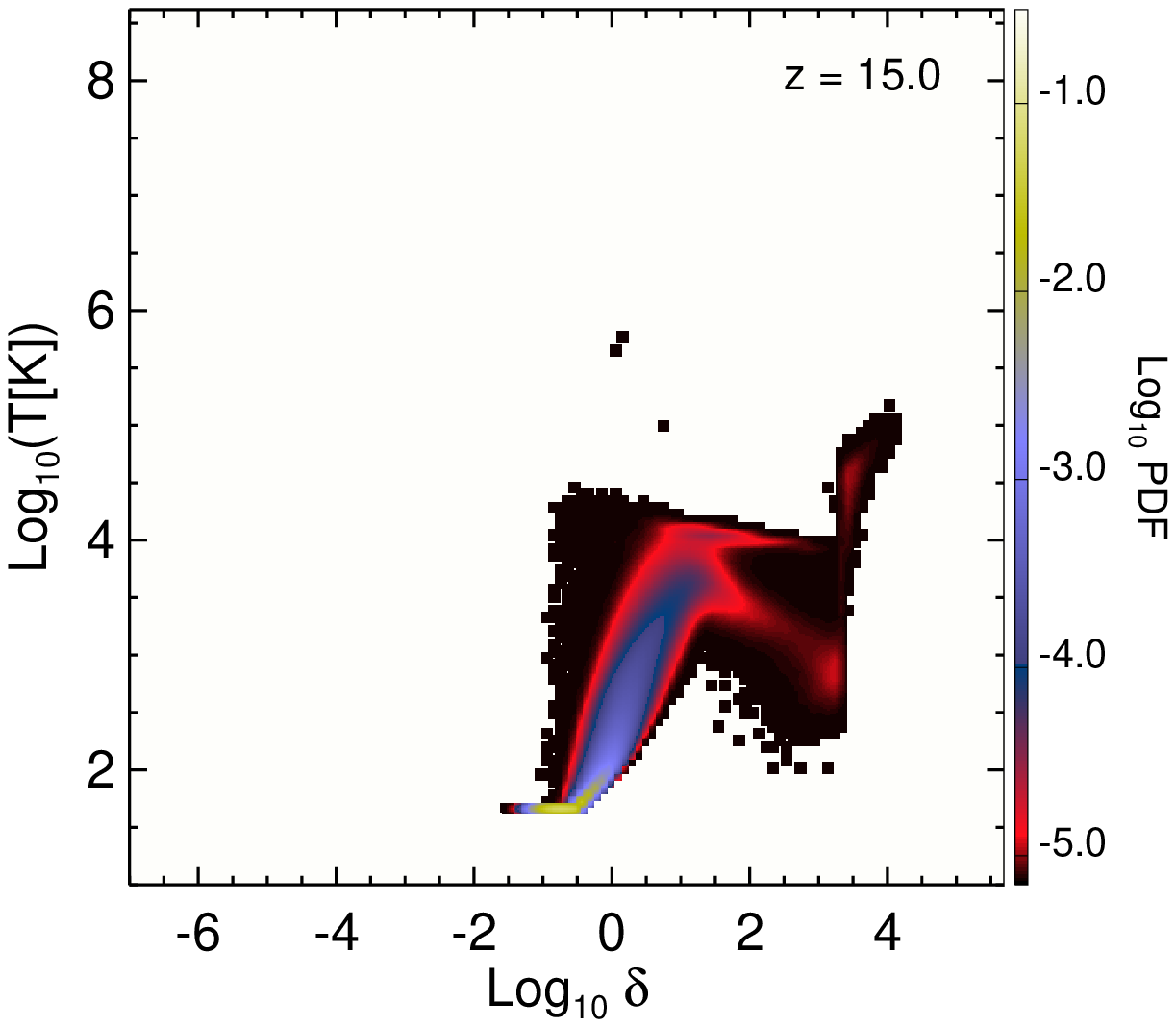}
  \hspace{-11.2cm}
  \includegraphics[width=0.35\textwidth]{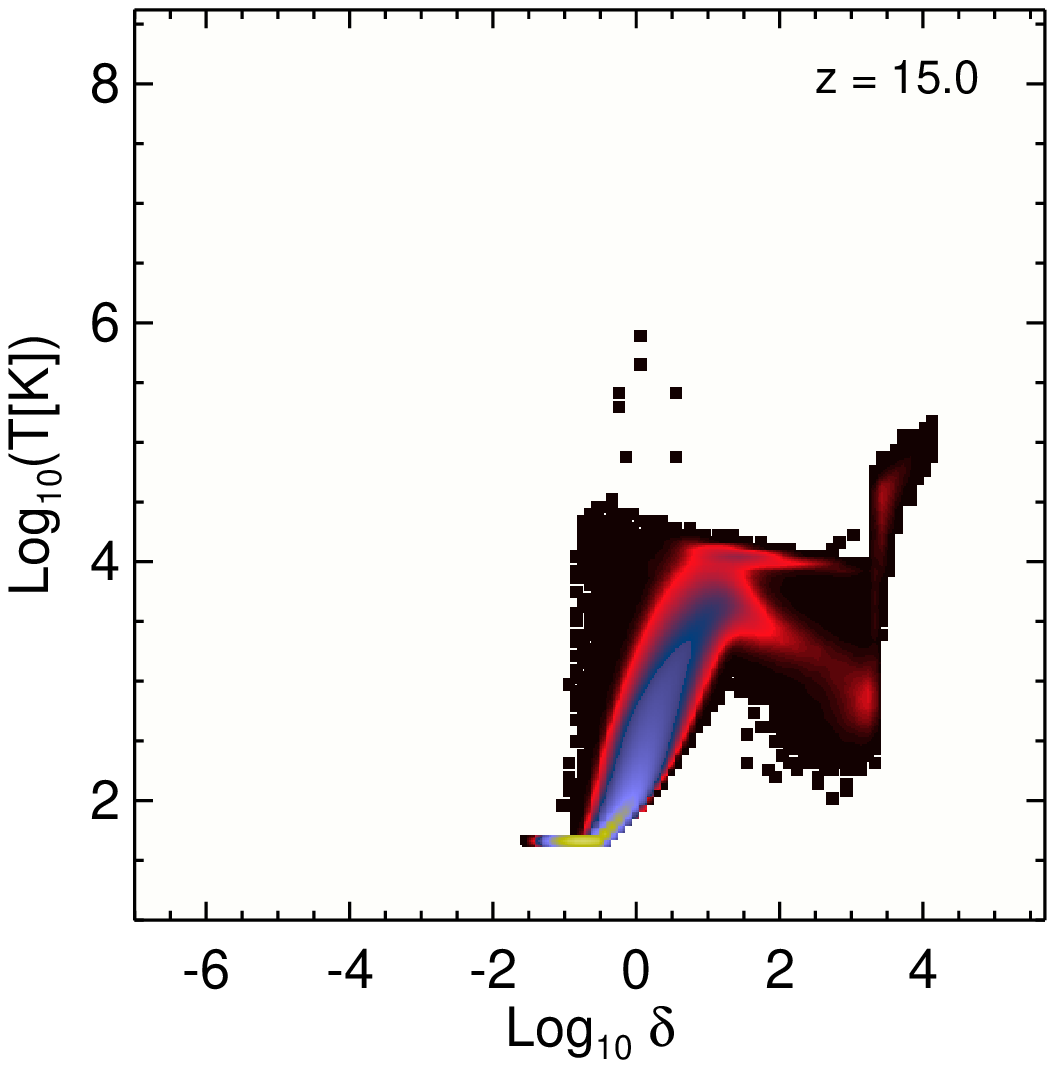}
  \hspace{-11.2cm}
  \includegraphics[width=0.35\textwidth]{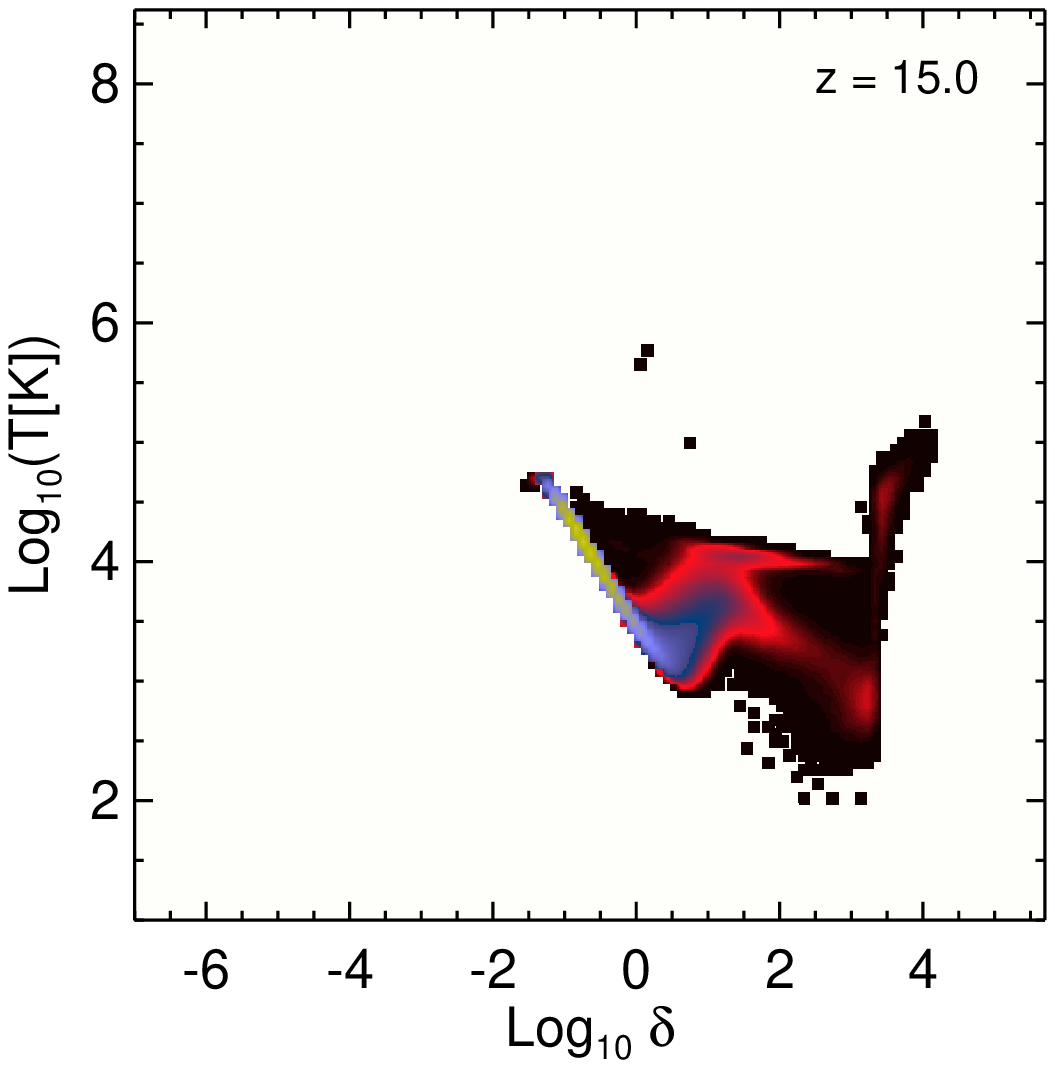} \\
    \vspace{-1.6cm}
  \includegraphics[width=0.35\textwidth]{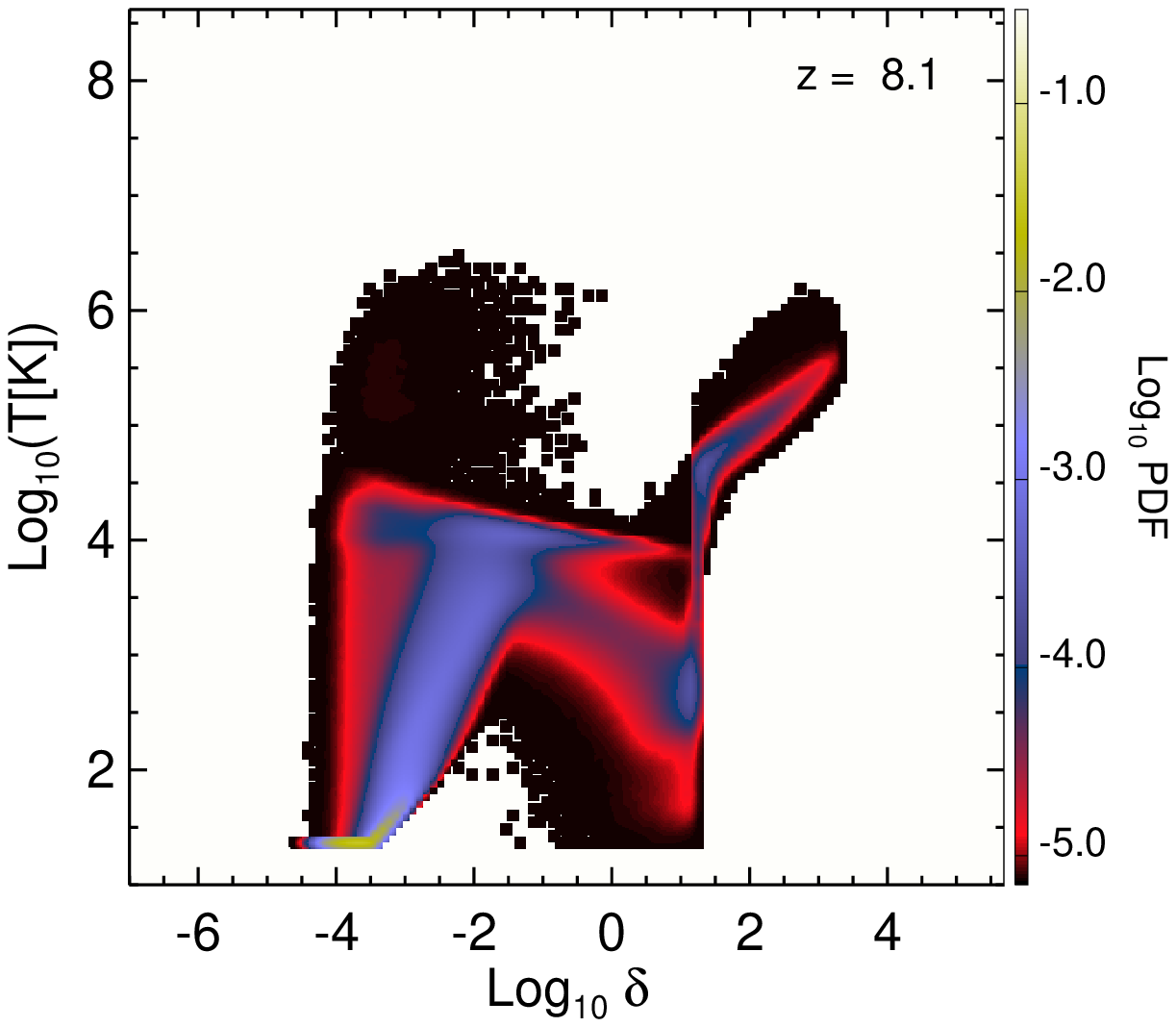}
  \hspace{-11.2cm}
  \includegraphics[width=0.35\textwidth]{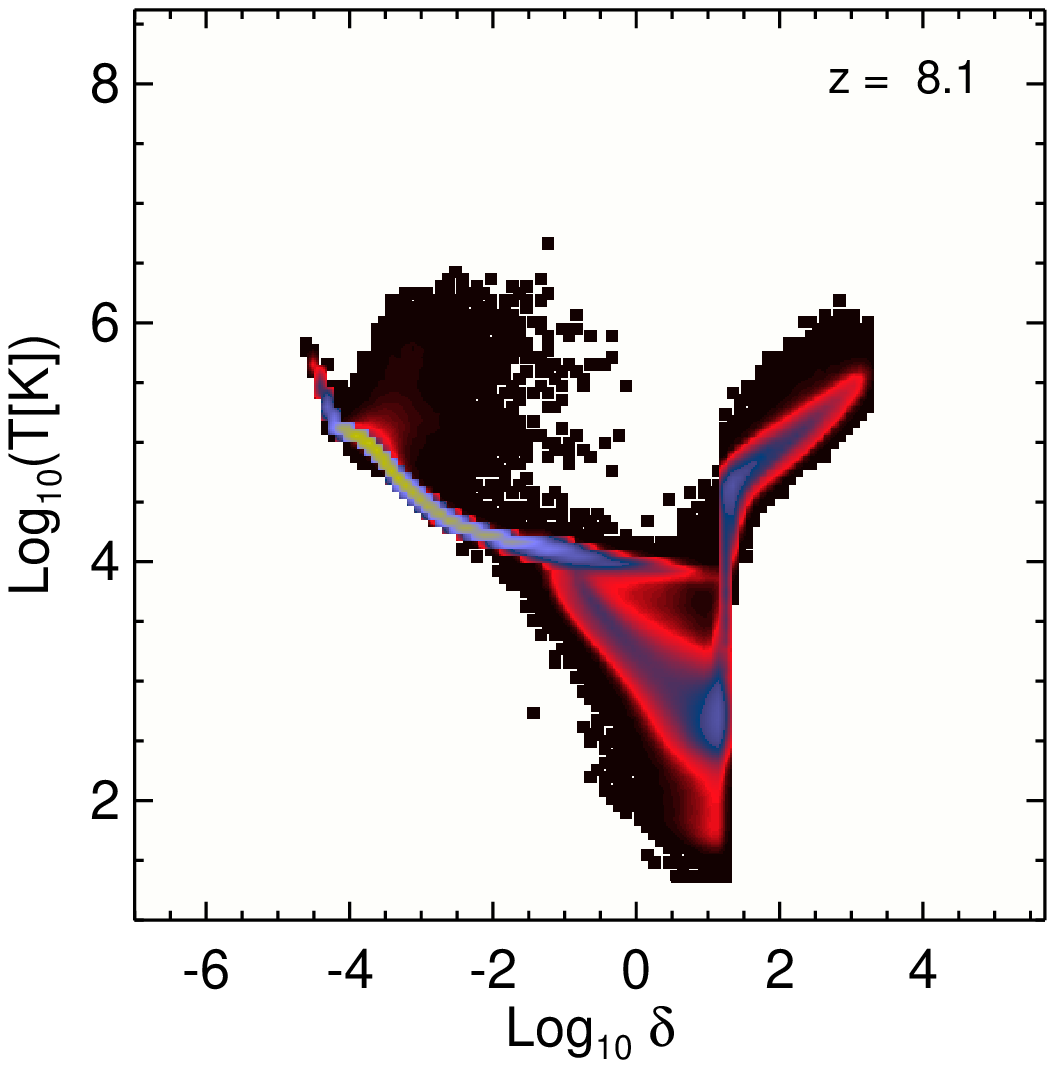}
  \hspace{-11.2cm}
  \includegraphics[width=0.35\textwidth]{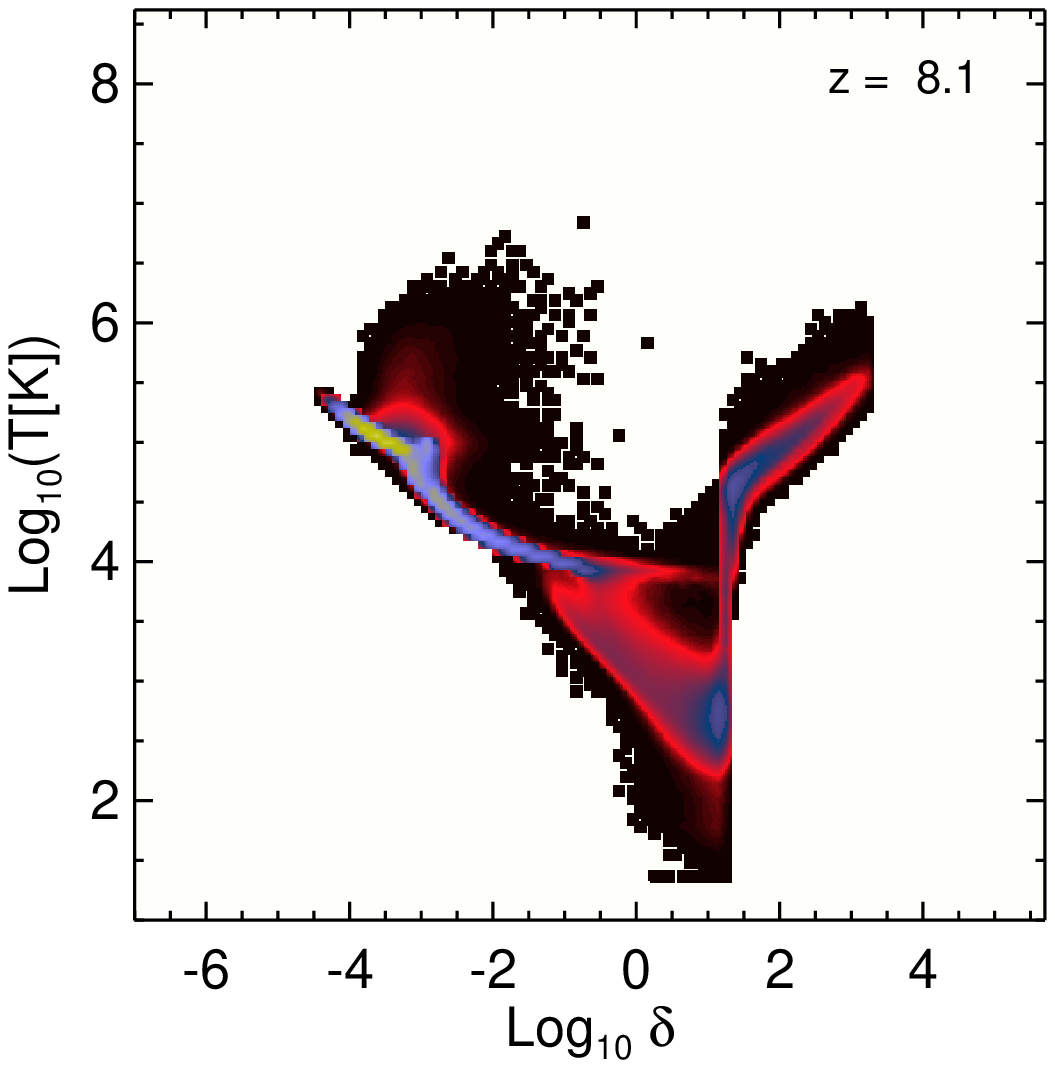}\\
\vspace{-0.4cm}
\caption[]{\small
Phase diagrams at redshift $z=15.0$ (upper row) and $z=8.1$ (lower row) for the simulations in Tab.~\ref{tab:properties} run with UV rates by
\cite{P2019} (P19, left), 
\cite{FG2020} (FG20, center) and 
\cite{HM1996} (HM, right).
The simulation data for temperature, $T$, and gas overdensity with respect to the gas mean, $\delta$, have been gridded on a base-10 logarithmic scale. The color coding refers to the total probability density function (PDF).
  }
  \label{fig:phase}
\end{figure*}
\begin{figure*}
  \small{
    \hspace{2.3cm} 0.01, 350 km/s 
    \hspace{2.8cm} 1, 350 km/s
    \hspace{2.8cm} 0.1, 700 km/s 
    }\\
  \vspace{-0.9cm}\\
\centering
\includegraphics[width=0.35\textwidth]{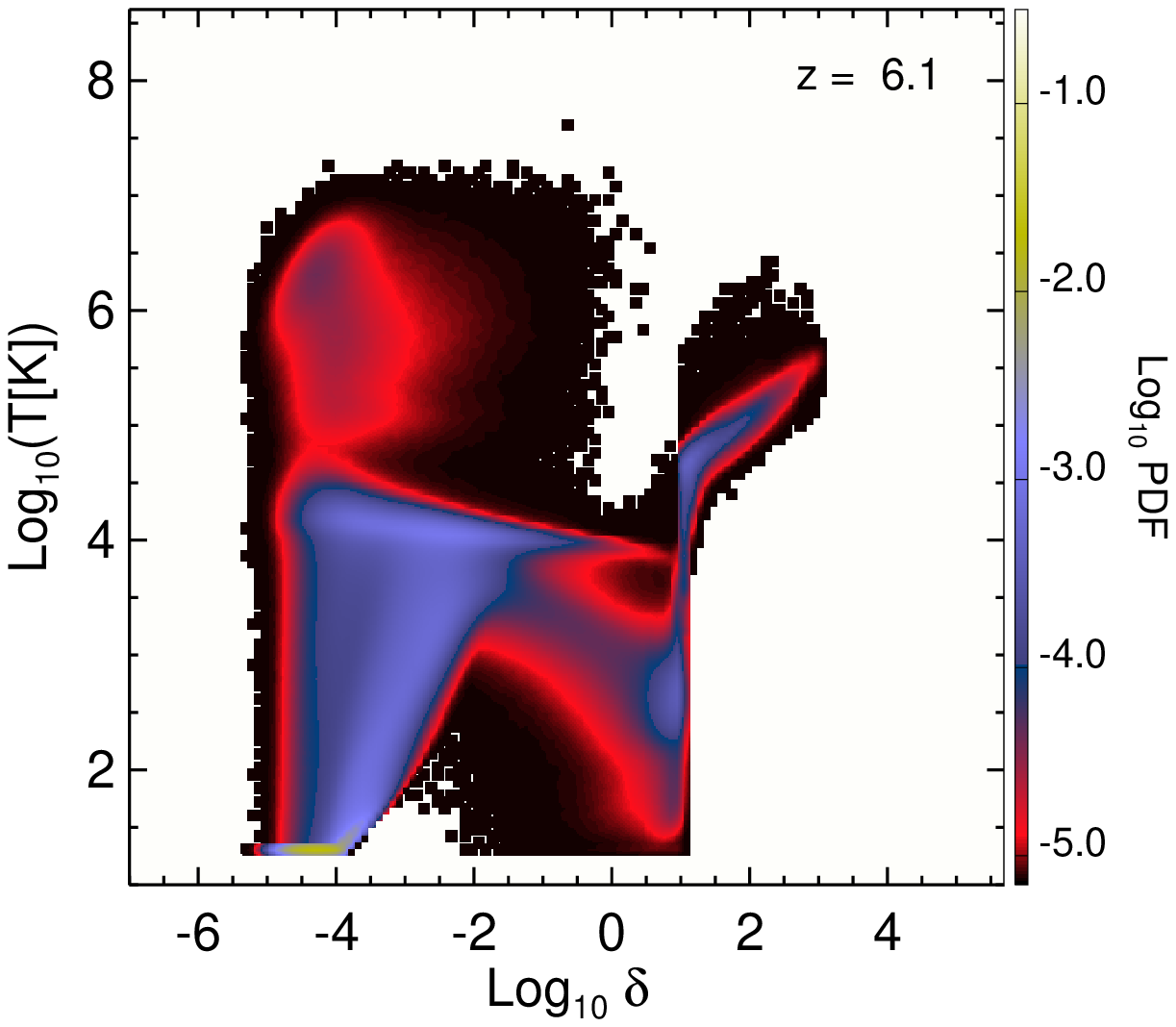}
 \hspace{-11.2cm}
\includegraphics[width=0.35\textwidth]{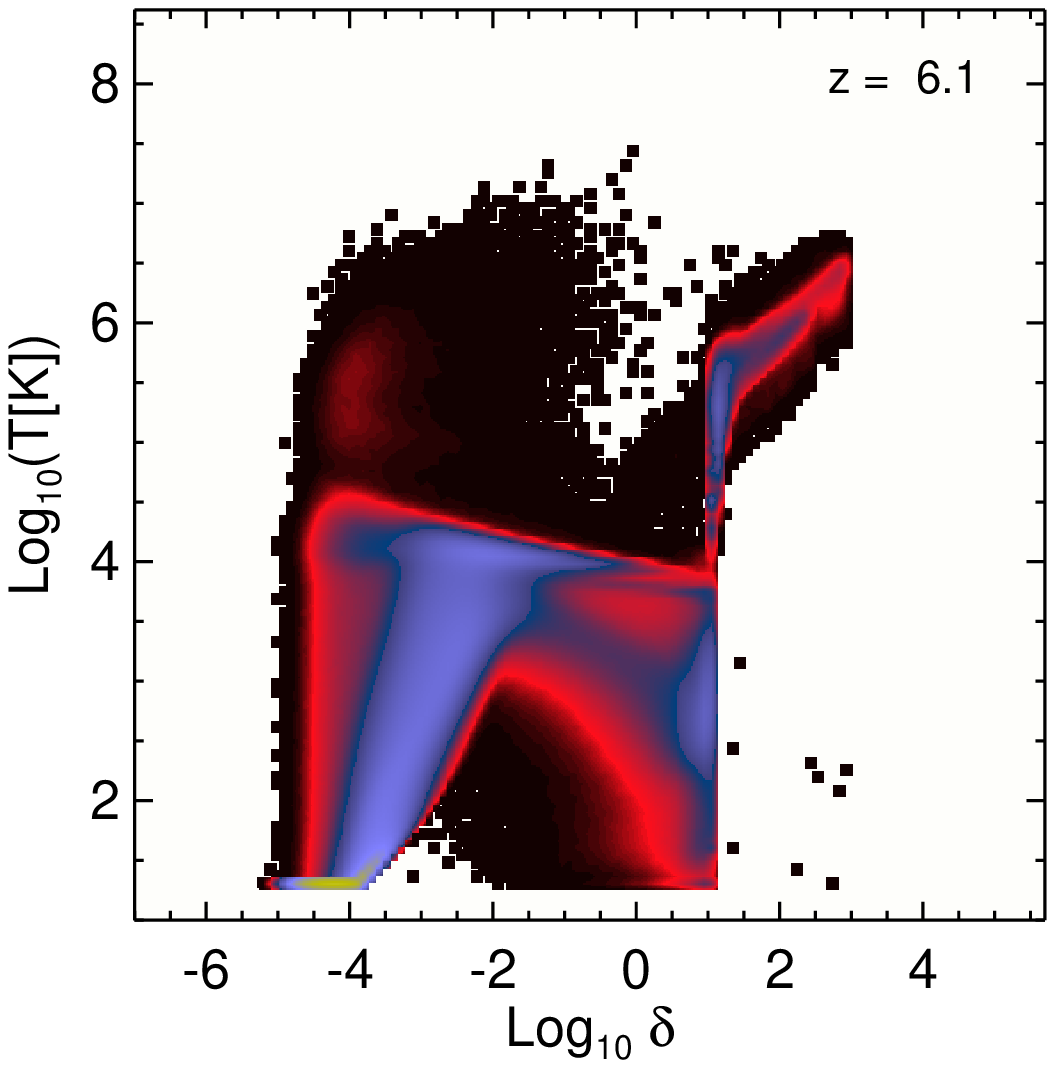}
 \hspace{-11.2cm}
\includegraphics[width=0.35\textwidth]{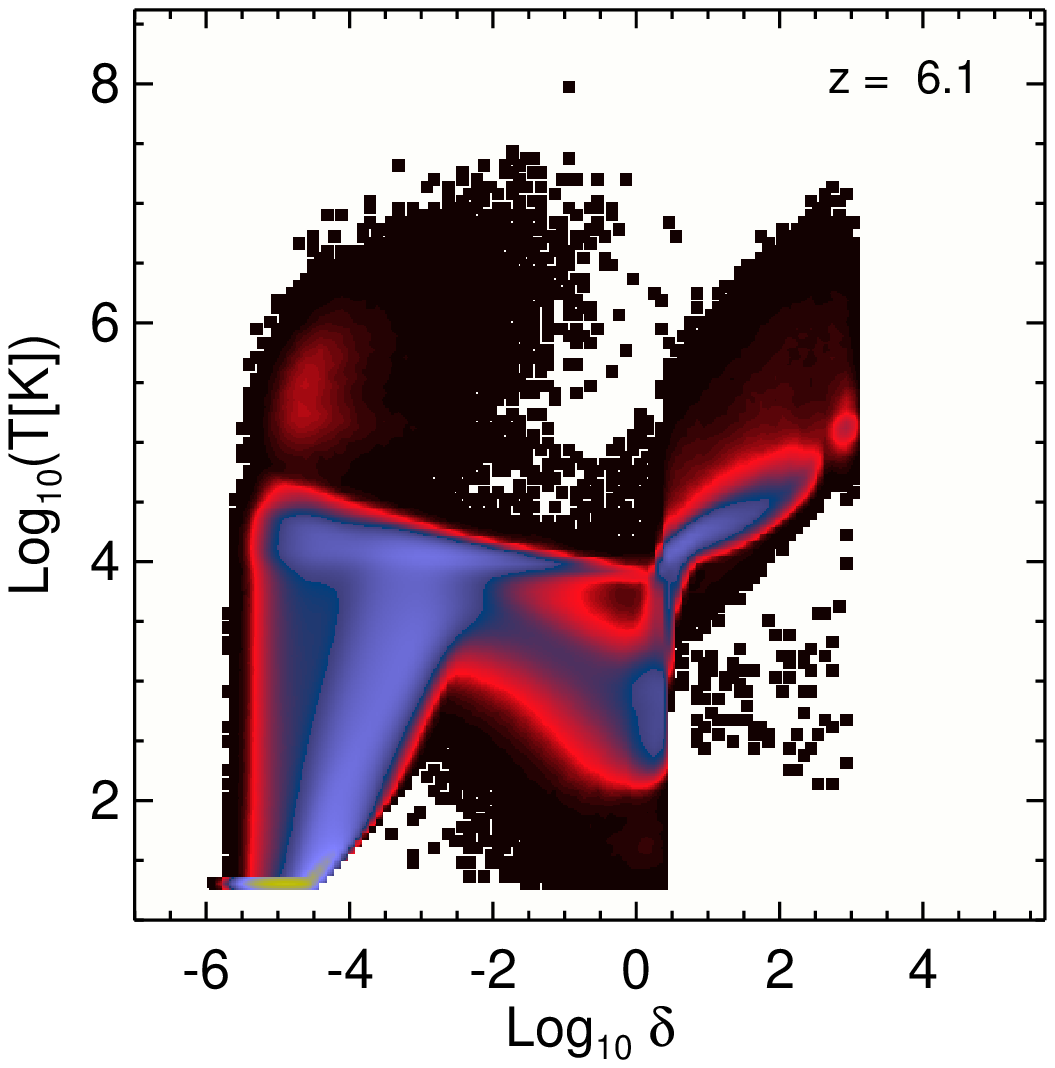}
\vspace{-0.4cm}
\caption[]{\small
Phase diagrams at redshift $z = 6.1$ in scenarios with HM background and, from left to right a 
SN efficiency (wind velocity) of:
0.01 (350 km/s), 
1 (350 km/s),
0.1 (700 km/s).
These are the runs HM-0.01, HM-1 and HM-w700, respectively.
}
\label{fig:phaseFeedback}
\end{figure*}
We check the impact of the different UV backgrounds by looking at the effects on the thermal state of cosmic gas.
In Fig.~\ref{fig:phase} we show the density-temperature diagram (phase space) for indicative runs, in terms of gas overdensity, $\delta = \rho / \overline{\rho}$,
where $\rho$ is the mass density of each gas parcel and $\overline{\rho}$ the mean gas density.
More specifically, phase space diagrams for three representative runs with 
P19, FG20 and HM UV rates, are displayed and the color code reflects the corresponding total probability distribution function.
In these models UV photoionization is switched on at $ z = $ 15.1 (P19), 8.3 (FG20) and 6 (HM).
At early times ($ z=15 $), trends in simulations run with FG20 and HM are similar, with most of the gas being cold, below $ 10^4 \,\rm K$.
This results from the natural evolution of gas chemistry in absence of UV radiation.
Low-density gas is shock-heated and the molecular fraction stays close to the initial-condition value (a few $ 10^{-6} $).
At higher densities, instead, non-equilibrium chemistry induces a rapid formation of H$_2$ molecules and consequent runaway cooling.
In the P19 case, IGM/CGM-like material (around and below the mean density) has just been warmed up by the newly injected UV photoheating and experiences an inversion of its equation of state.
High-density regions are little affected, though, as a result of gas clumpiness in such regimes.
Obviously, how photoheating and photodissociation proceed depend on the chosen scenario, as highlighted by the lower panels at $z=8.1$,
when the gas thermal evolution has reached different stages:
the traditional HM background has not heated cold, thin, cosmic gas, yet;
in the FG20 and P19 panels, instead, photoheating is clearly effective since
$z=8.3$ for FG20 and $z=15.1$ for P19.
We note that the runaway molecular cooling branch is only mildly affected by UV radiation, as a consequence of the larger densities that shield H$_2$ molecules.
At $z\lesssim 6$ all three models present phase diagrams similar to the latter case.
We note that, irrespectively from the chosen UV background rates, high-density regions behave in a similar way, dictated by local thermal processes rather than background radiation. 
Indeed, the star-forming branch exhibits the equation of state of SN-heated gas based on the multiphase ISM model of \cite{SpringelHernquist2003}. 
This behaviour is caused by the SN energy released in the surrounding medium and is responsible for the increase in gas temperatures visible on the right of each the panel. \\ 

\noindent
Collapsing star forming regions are more sensitive to feedback effects and the implications of different assumptions for feedback parameters might be relevant for the thermodynamical evolution of cosmic gas, independently from the adopted UV prescriptions.
As indicative examples, in Fig.~\ref{fig:phaseFeedback} we show phase diagrams at redshift $z = 6.1$ for the runs with SN efficiencies of 0.01-1 and wind velocity of 350 or 700~km/s in the HM case.
A low SN efficiency of 0.01 (left panel, HM-0.01) is incapable of removing dense material from the star forming regions (at $\rm Log_{10}\delta \gtrsim 1 $), where gas retains temperatures of $10^4$-$10^5\,\rm K$ without cooling further (red/blue branch).
A larger efficiency of 1 (central panel, HM-1) removes more gas, as clear from the probability distribution of the star forming branch (red region at $\rm Log_{10}\delta \sim 1$-2.5). Such material circulates through the IGM/CGM again and becomes available for further cooling and fragmentation.
The 0.1 case with the same winds (not plotted) features an intermediate behaviour.
The results displayed both in the left and in the central panel have a wind velocity of 350~km/s.
Stronger winds with velocities of e.g. 700~km/s (right panel with SN efficiency of 0.1, HM-w700) impact gas removal from star forming sites. 
In the case shown, dense material is ejected to outer regions more efficiently and populates abundantly the thin IGM at  $\rm Log_{10} \delta \lesssim -1$ and  $T \gtrsim 10^5\,\rm K$.
Besides the role of wind velocity, other wind parameters, such as the wind delay time before re-coupling to hydrodynamics (e.g. $0.1 \, t_{\rm H}$ adopted in HM-wf, instead of the reference $0.025 \, t_{\rm H}$ adopted in Fig.~\ref{fig:phaseFeedback}) are expected to have less vital effects.
The picture at different redshifts is qualitatively similar, as it is when assuming a Chabrier IMF (HM-Chabrier and HM-1-Chabrier), instead of a standard Salpeter IMF. Alternative UV rates for these models (e.g. P19, P19-1, P19-Chabrier, P19-1-Chabrier) show analogous behaviours, except for the mentioned (temporal) differences.
These preliminary considerations highlight the main impact of different assumptions on the UV background and feedback effects, specifically on the thin cosmic medium.
The implications for cold dense media (privileged sites hosting neutral and molecular gas) are analyzed in the next Sect.~\ref{Sect:Omega}.

\subsection{Density parameters}
\label{Sect:Omega}
\begin{figure}
\includegraphics[width=0.49\textwidth]{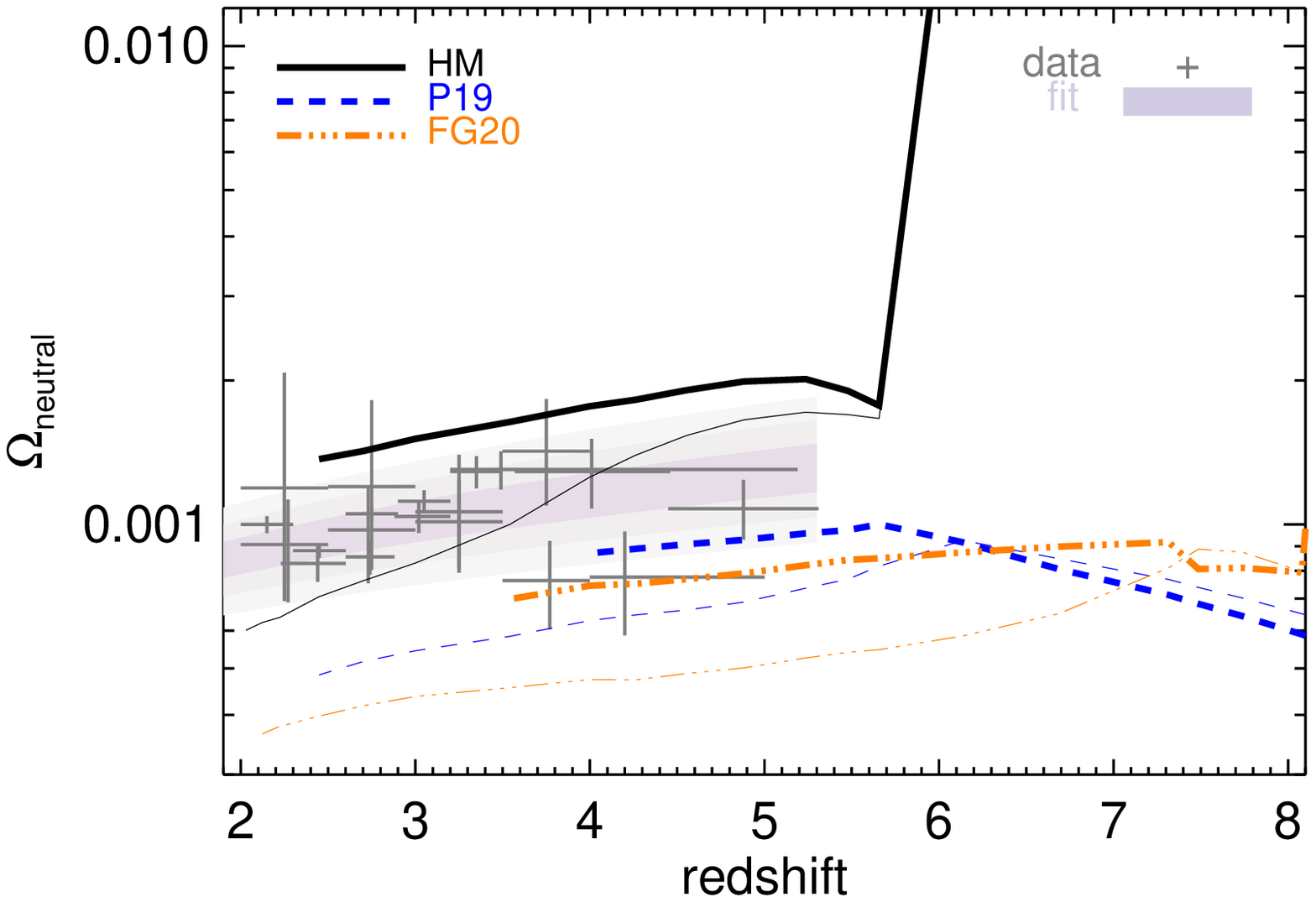} \\
\includegraphics[width=0.49\textwidth]{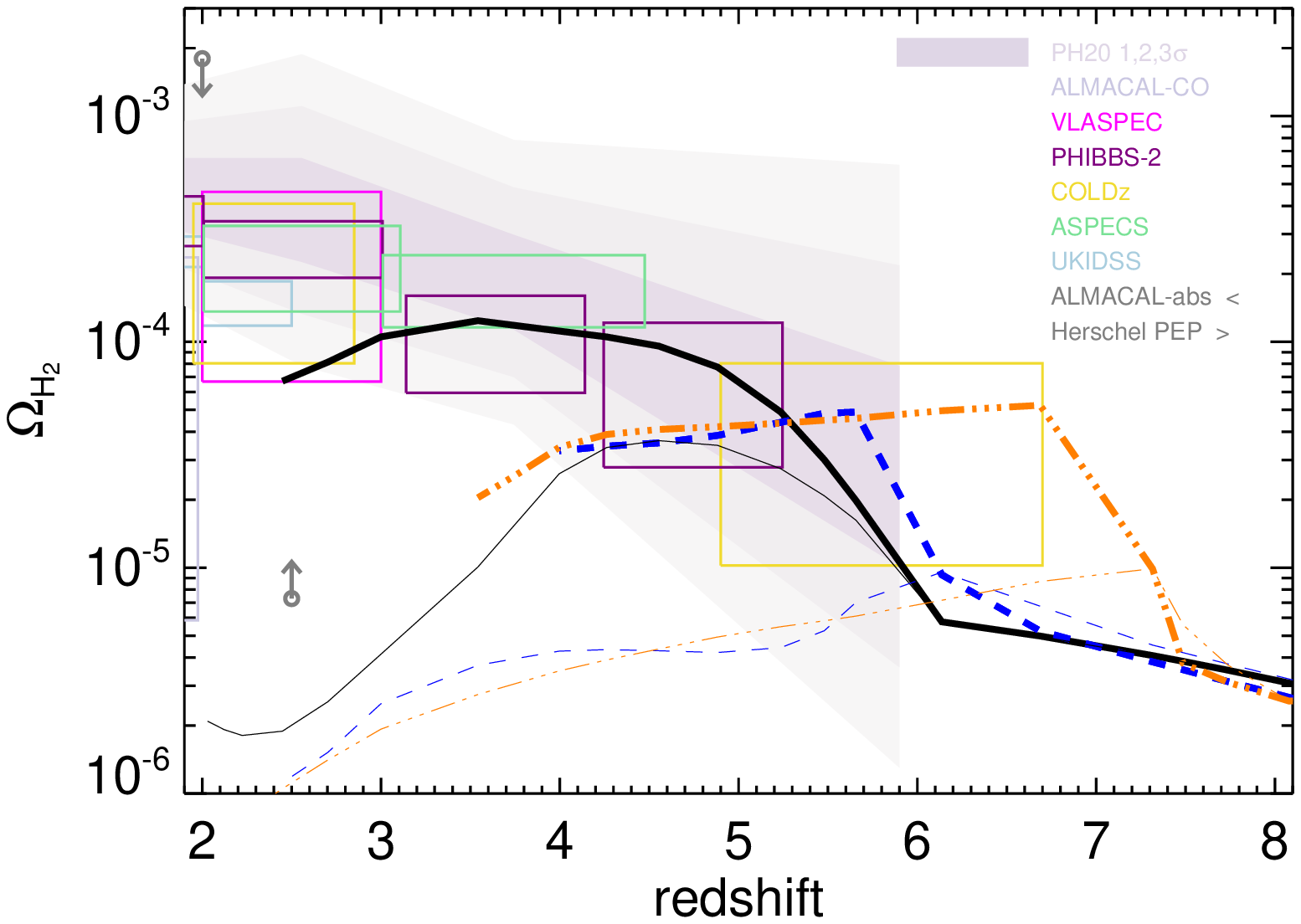}
\caption[]{\small
Upper panel:
redshift evolution of $ \Omega_{\rm neutral} $ compared to inferred values from HI observations with $1\sigma$ error bars (grey points) 
and corresponding fit (from darker to lighter shades) at 1, 2 and 3$\sigma$ confidence levels \cite[see][]{PerouxHowk2020}.
The simulations considered are:
HM-HISSmed (thick black solid lines),
P19-HISSmed (thick blue dashed lines)
and FG20-HISSmed (thick orange dot-dot-dot-dashed lines).
They all include HI self-shielding.
As a comparison, corresponding un-shielded runs with the same UV backgrounds (HM, P19 and FG20) are displayed by thin lines of the same color and style.
Lower panel:
corresponding evolution of $\Omega_{\rm H_2}$ compared to the observational constraints listed in Tab.~\ref{tab:dataOmegaH2}
(PH20 1, 2 and 3$\sigma$ confidence levels, shades;
ALMACAL-CO, light grey;
VLASPEC, magenta;
PHIBBS-2, purple;
COLDz, yellow;
ASPECS, green;
UKIDSS-UDS, cyan;
ALMACAL-abs and Hershel PEP 
limits, dark grey).
}
\label{fig:OmegaUVdata}
\end{figure}

\noindent
In the following we focus on the evolution of mass density parameters for neutral gas, $\Omega_{\rm neutral} $, 
and molecular H$_2$ gas, $ \Omega_{\rm H_2} $,
as extracted from simulation outputs at different times,
and compare them to state-of-the-art observational data 
(Sect. \ref{sect:data}) 
in Fig.~\ref{fig:OmegaUVdata}.
Theoretical trends refer to runs with the reference implementation including time-dependent non-equilibrium abundance computations,
a high-density threshold of $ 10\,\rm cm^{-3} $ for 
stochastic star formation,
stellar evolution,
metal spreading,
and
H$_2$ self-shielding.
We distinguish between simulations accounting for HI self-shielding, 
(HM-HISSmed, P19-HISSmed and FG20-HISSmed denoted by thick lines)
and analogous un-shielded results (HM, P19 and FG20 denoted by thin lines).
\\

\noindent
In the top panel, we focus on $ \Omega_{\rm neutral} $.
We can see that theoretical results increase smoothly with redshift at $z \simeq 2$-5, in broad agreement with observational data.
UV radiation impacts our results after the on-set of the UV background and leaves the signal from earlier epochs unaffected (as visible for all the UV models e.g. in Appendix~\ref{AppendixOmegaUV}).
In both HI shielded and un-shielded HM scenarios, the cold mass surviving at $ z \lesssim 6$, after photoionization is turned on, is only 5\% of the original one, corresponding to $\Omega_{\rm neutral} \sim 10^{-3}$. 
As HI shielding preserves HI mass, the resulting $\Omega_{\rm neutral} $ at $z\simeq 2$-3 is larger by a factor of up to 2 with respect to the un-shielded case.
Although both curves are broadly consistent with the available data at $ z \lesssim 5$, the un-shielded run seems to predict a steeper evolution of cold gas.
The shielded HM scenario is slightly above the 1$\sigma$ error bars of observational data points.
In the P19 and FG20 cases the UV model is turned on earlier, resulting in a larger effect of the associated feedback and flatter curves (compared to those obtained with an HM background).
In these cases, the predictions are slightly below 1$\sigma$ observational determinations for the un-shielded runs, while the corresponding shielded cases are roughly consistent with observational lower limits.
The addition of further physical processes, as we will see in the following, does not strongly affect $\Omega_{\rm neutral}$, whose values remain robustly close to the ones displayed in the figure and within the error bars of available data.
We note that the displayed fitting expression by \cite{PerouxHowk2020}, 
i.e. $ \Omega_{\rm neutral } \propto (1+z)^{0.57} $,
is similar to other independent formulations finding
$ \Omega_{\rm neutral } \propto (1+z)^{0.60} $ 
\cite[][]{Crighton2015} 
or 
$ \Omega_{\rm neutral } \propto (1+z)^{0.64} $
\cite[][]{Rao2017}.
Thus, uncertainties on the numerical fit do not affect our conclusions.
\\

\noindent
In the bottom panel of Fig.~\ref{fig:OmegaUVdata} we explore the outcome for $\Omega_{\rm H_2}$, compared to the observational constraints (listed in Tab.~\ref{tab:dataOmegaH2}) by
\cite{Hamanowicz2021} (ALMACAL-CO), 
\cite{Riechers2020vlaspec} (VLASPEC),
\cite{Lenkic2020} (PHIBBS-2) 
\cite{Riechers2020coldz} (COLDz),
\cite{Decarli2020} (ASPECS),
and
\cite{Garratt2021} (UKIDSS-UDS).
For ALMACAL-CO maximum and minimum values of preliminary estimates at $z<2$ are shown.
Darker to lighter shades correspond to 1, 2 and 3$\sigma$ confidence levels by \cite{PerouxHowk2020}.
ALMACAL-abs upper limit in the $z\simeq 0$-2 range is taken by \cite{Klitsch2019} while Herschel PEP most probable lower limit at z$\simeq2-$3 is by \cite{Berta2013}.
\\
H$_2$ mass forms in cold high-density regions with rates depending on gas temperature and ionization state.
Therefore, as opposed to $\Omega_{\rm neutral}$, predicted values for $\Omega_{\rm H_2}$ are very sensitive to the underlying chemical and physical modelling.
We neatly see that HI shielding effects in the HM-HISSmed, P19-HISSmed and FG20-HISSmed runs
are responsible for largely increasing the amount of H$_2$  with respect to the corresponding un-shielded HM, P19 and FG20 results.
Indeed, the $\Omega_{\rm H_2}$ parameter is enhanced by $\sim $1 dex, while the HI mass increases only by a factor up to 2.
This reveals the highly non-linear nature of the involved chemical processes, as to a change of a factor of a few in HI correspond much larger variations in H$_2$.
Clearly, the un-shielded HM, P19 and FG20 models produce excessive H$_2$ dissociation and are not consistent with the data (the related $ \Omega_{\rm H_2} $ values fall down to $\lesssim 2 \times  10^{-6}$ at $z \lesssim 3 $, far from observational limits by almost to 2 orders of magnitude).
Despite the inclusion of HI shielding, though, the P19 and FG20 scenarios feature an $ \Omega_{\rm H_2} $ evolution that flattens at $z \simeq 4$-7, hardly reaching $ \Omega_{\rm H_2} \sim 10^{-4} $.
This is a result of the earlier onset of the UV background and more powerful rates associated to these two models at such epochs.
Conversely, the HI-shielded HM model produces values closer to the observed ones by employing only the essential H$_2$ formation channels, which are able to enhance the H$_2$ mass density parameter at $z \simeq 4$-6 in the presence of a milder UV background.
Indeed, the evolution of the H$_2$ density parameter is affected by the evolution of the photoionization and photoheating rates adopted in the simulations.
The P19 rates evolve smoothly from high redshift until $z\simeq 6$, when they get enhanced by more than 3 orders of magnitude, while the FG20 rates at $z\simeq 8$ go from zero to levels comparable to those of the P19 model.
Although newly available free charges induce H$_2$ formation, on longer timescales the overall effect is H$_2$ disruption, in shielded as well as un-shielded cases.
The decline observed in $\Omega_{\rm H_2}$ at lower redshift is linked to the stronger effect of both UV radiation and star formation, which destroy molecules through photoheating and stellar feedback.
The latter is increasingly effective at later times (see e.g. lower-$z$ trends for un-shielded models), while the former depends on the details of the UV background at different $z$.
For an earlier and/or stronger UV background (e.g. P19 and FG20) H$_2$ suppression is more significant, albeit modulated by gas self-shielding.
\\
By looking at $\Omega_{\rm H_2}$ data in light of our findings, one might think that the large dispersion in the observations at low redshifts 
\cite[e.g.][]{Klitsch2019, Decarli2019, Decarli2020, Riechers2020coldz, Garratt2021} 
could be related to targeted objects found in conditions that are heterogeneous in terms of gas shielding, heating, ionization or photodissociating radiation (hence, the need for large statistical samples).
In practice, the right combination of UV model and HI self-shielding is needed to match both HI and H$_2$ data.
A UV background turning on at $ z \gtrsim 8 $ seems to induce an excessive feedback and predict too little cold gas.
This suggests that either UV rates at those times should be lower than those adopted here or that high-$z$ gas self-shielding should be stronger.
\\

\begin{figure}
\includegraphics[width=0.49\textwidth]{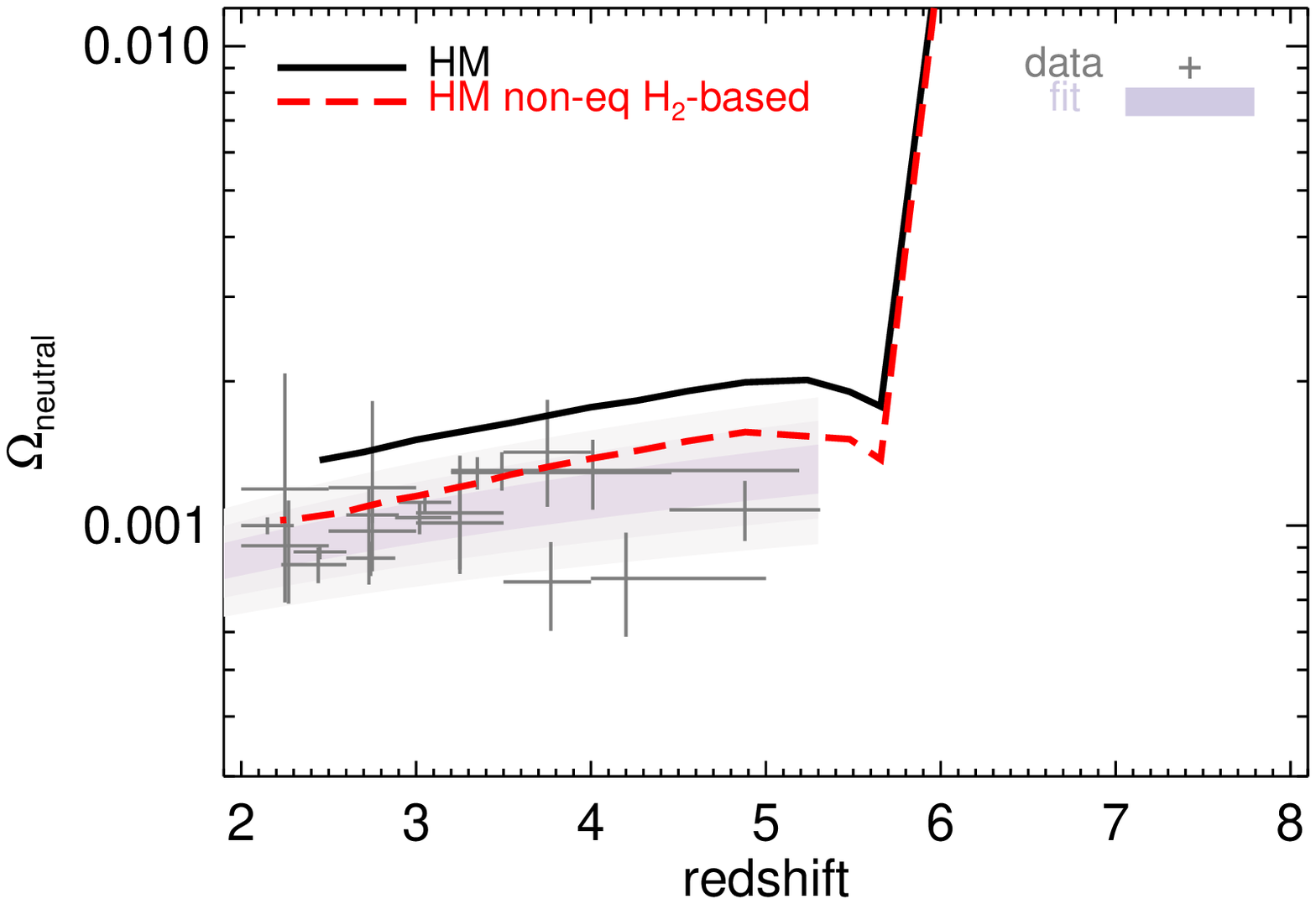}\\
\includegraphics[width=0.49\textwidth]{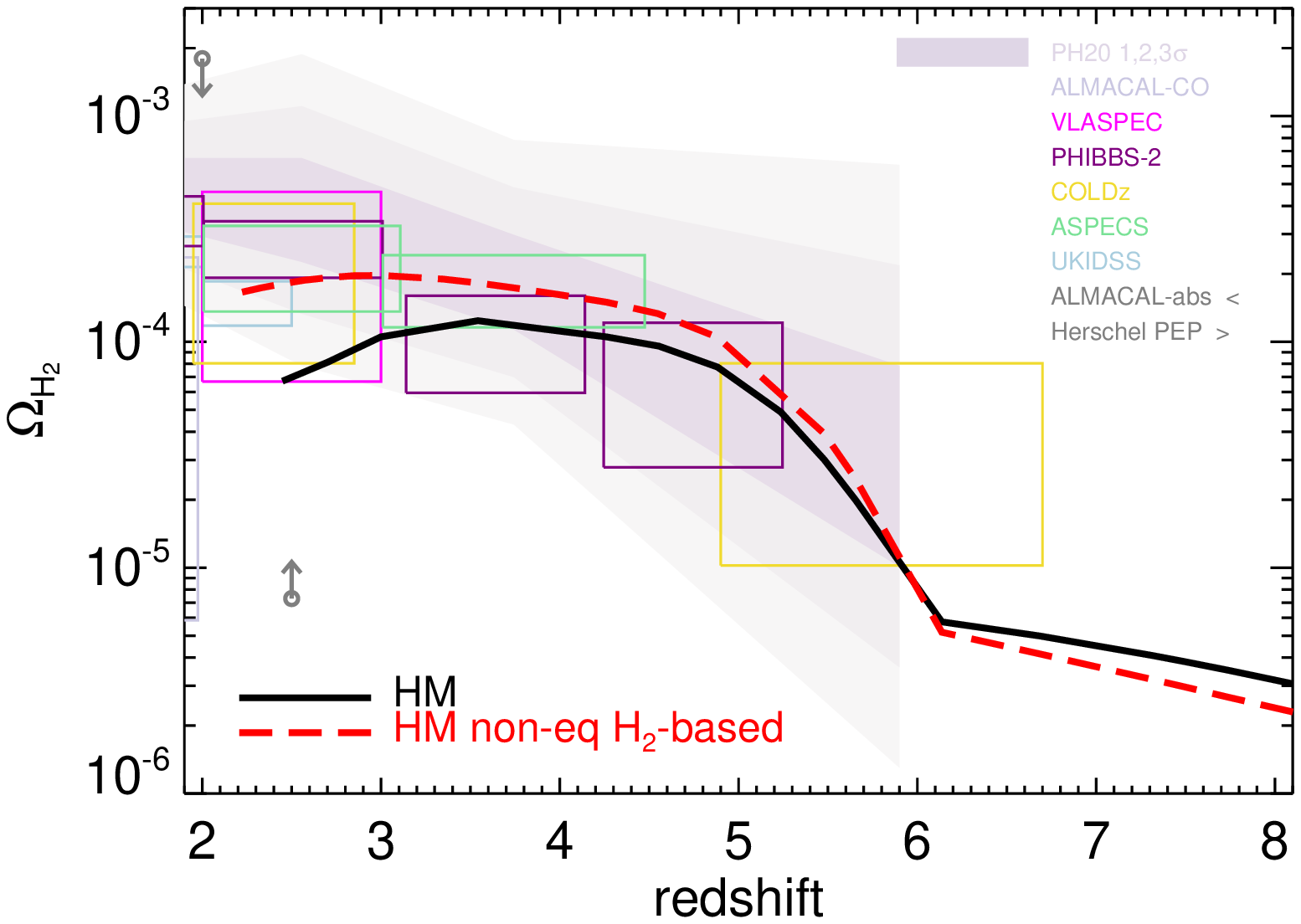}
\caption[]{\small
Upper panel:
$\Omega_{\rm neutral}$ evolution for the non-equilibrium H$_2$-based star formation model (red dashed line, HM-HISSmed-H2) with an HM background and compared to the reference run (solid black line, HM-HISSmed).
Lower panel:
Corresponding $\Omega_{\rm H_2}$ evolution.
Observational determinations are listed in Tab.~\ref{tab:dataOmegaH2} (see also text and Fig.~\ref{fig:OmegaUVdata}).
}
\label{fig:OmegaUVdata_full}
\end{figure}

\noindent
Typically, stochastic conversion of gas into stars is responsible for removing very dense gas parcels from collapsing material.
Nevertheless, $ \Omega_{\rm neutral} $ and $ \Omega_{\rm H_2} $ parameters might be affected if gas-to-stars conversion involves particles that are still at relatively low density (below $\sim$1 particle $\rm cm^{-3}$).
In this case the abundance of HI and H$_2$ cannot grow significantly during their cooling process.
To make sure that gas evolution through the cooling branch is followed properly and the amount of HI and H$_2$ formed is realistic, we have always used a density threshold of 10~cm$^{-3}$, as time-dependent non-equilibrium runaway cooling usually takes place in gas with densities $ > 0.1 \,\rm cm^{-3}$  and temperatures $\sim 10^2-10^4 \,\rm K$.
Besides the density threshold, though, gas-to-star conversion is often assumed to depend on gas density, $\rho $, and a typical star formation timescale, $t_{\rm sf} $ (see Sect.~\ref{SFmodel}).
This means that the particle SFR is $\propto \rho / t_{\rm sf}$.
Since H$_2$ molecules are the main drivers of star formation (not the gas {\it tout court}) and since we are able to follow their non-equilibrium chemical abundance precisely, in the following we employ the local H$_2$ density, $\rho_{\rm H_2}$, as derived by our non-equilibrium calculations to estimate the particle SFR  $\propto \rho_{\rm H_2} / t_{\rm sf}$ and related feedback effects (non-equilibrium H$_2$-based star formation).
In Fig.~\ref{fig:OmegaUVdata_full} we compare the resulting $ \Omega_{\rm neutral} $ and $ \Omega_{\rm H_2} $ parameters for such scenario (HM-HISSmed-H2 run) to the reference results (HM-HISSmed run).
Although the evolution of $\Omega_{\rm neutral}$ and $\Omega_{\rm H_2}$ obtained with the H$_2$-based approach is similar to the reference one, it seems to be a better fit to observations of both gas components.
With respect to the reference trends, $\Omega_{\rm H_2}$ values are slightly higher as consequence of the less effective star formation feedback. Consequently, HI is more easily converted into H$_2$ and resulting $\Omega_{\rm neutral}$ values are smaller.\\

\noindent
We have additionally investigated the role of individual H$_2$ formation channels, finding that H$^-$ is a fundamental driver at all times, while three-body interactions are less important on large scales.
Nevertheless, they are useful to reach large H$_2$ fractions of at least $50\%$ (i.e. $\sim 67\%$ of the available H mass) even during the epoch of reionization and explain the $50-60\%$ values at $z\sim 4$-6 lately reported by e.g. 
\cite{DZ2020} and \cite{Tacconi2018, Tacconi2020}
(Appendix \ref{AppendixH2channels}).
Including gas-grain processes coupled to the non-equilibrium molecular network and metal spreading leads to a more complete treatment of cold-gas chemistry, without substantially affecting our conclusions.
Also dust grains can enhance H$_2$ fractions up to $\sim 50\% $ by $z\simeq 7$, in line with observations (for a more detailed discussion see Appendix \ref{AppendixH2gr} and \ref{AppendixPE}), but are effective at relatively large metallicities.
Heating due to cosmic rays formed in star forming regions is rather localized and, no matter the details of the ionization rates, its impact remains at a ten-per-cent level (Appendix \ref{AppendixZeta}).
Feedback due to different SN efficiencies or wind parameterizations impact $\Omega_{\rm neutral}$ and $\Omega_{\rm H_2}$, inhibiting more or less efficiently the amount of cold neutral and molecular gas available.
Material in the distant IGM/CGM suffers only mild effects from these events (Appendix \ref{AppendixSN} and \ref{AppendixWindsIMF}).
\\
To summarize, we find that to model accurately cold gas and to reproduce observed atomic and molecular abundances it is crucial to follow non-equilibrium chemistry of HI and H$_2$ self-shielded material down to densities of at least $ \sim 10\,\rm cm^{-3}$ and to resolve the gas runaway cooling.
In this way, our results are in better agreement with observations than previous simulation predictions at high $ z $.

\subsection{Gas depletion times}

\begin{figure}
\includegraphics[width=0.49\textwidth]{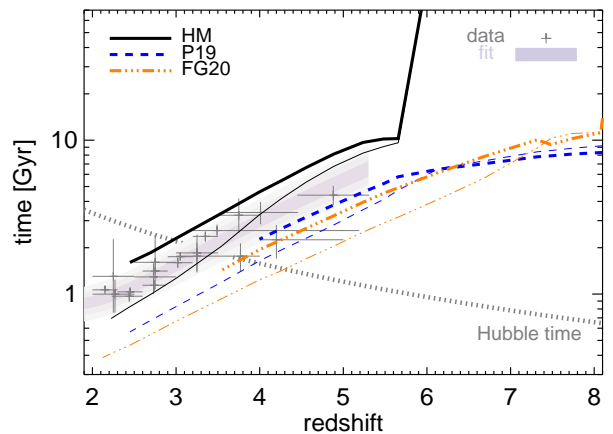}
\caption[]{\small
Redshift evolution of the neutral-gas depletion times in simulations with HM, P19 and FG20 (solid, dashed and dot-dot-dot-dashed lines, respectively) UV backgrounds compared to depletion times with $1\sigma$ errors  (grey points) derived from $ \Omega_{\rm neutral} $ observational data and 1, 2 and 3$\sigma$ confidence levels  (from darker to lighter shades) (see also Fig.~\ref{fig:OmegaUVdata}).
Results are for HI shielded (thick) and HI un-shielded (thin) scenarios.
The Hubble time (grey dotted line) is shown, as well.
}
\label{fig:tdepl_multicold_bySFRD_data}
\end{figure}

Gas depletion times are useful tools to investigate the baryon cycle and its relation to cosmic expansion (often parameterized by the Hubble time, $t_{\rm H}$).
As hot gas is unlikely to collapse, it is natural to focus on cold neutral gas and on molecular H$_2$-rich gas.
\\
Fig.~\ref{fig:tdepl_multicold_bySFRD_data} shows cold-neutral-gas depletion time for different UV models, compared to observational data and the Hubble time.
For all the UV models, depletion times decrease with decreasing redshift, suggesting that typical cosmic structures can accumulate cold material that will eventually collapse and form stars.
This qualitative behaviour is basically universal and little dependent on model details.
Quantitatively, results from simulations including HI shielding are more consistent with the data, with values becoming lower than $t_{\rm H}$ at $z\lesssim 4$. 
Un-shielded models with P19 and FG20 background radiation are only marginally off observational constraints.
\\
\begin{figure}
\includegraphics[width=0.49\textwidth]{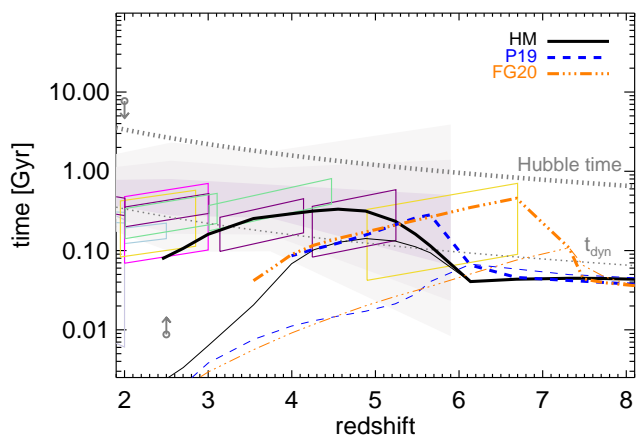}
\caption[]{\small
Redshift evolution of the H$_2$ depletion time in simulations with HM, P19 and FG20 (solid, dashed and dot-dot-dot-dashed lines, respectively) UV backgrounds.
H$_2$ depletion times within $1\sigma$ range (slanted shapes), 1, 2 and 3$\sigma$ confidence levels (from darker to lighter shades), as well as maximum and minimum values (grey arrows) are derived from $ \Omega_{\rm H_2} $ observational data and limits in Tab.~\ref{tab:dataOmegaH2} and Fig.~\ref{fig:OmegaUVdata}.
The Hubble time (thick grey dotted line) and the dynamical time, $t_{\rm dyn}$ (thin grey dotted line) are shown, as well.
Results refer to both HI shielded (thick) and HI un-shielded (thin) scenarios.
}
\label{fig:tdepl_multimol_data}
\end{figure}
Fig.~\ref{fig:tdepl_multimol_data} shows the corresponding H$_2$ depletion times compared to observational data.
In this case, besides the Hubble time, we also plot the `dynamical' time, $t_{\rm dyn}$, defined as a fraction of 10 per cent of the Hubble time, $ t_{\rm dyn} = 0.1\,t_{\rm H} $.
This expression can be explained with simple dynamical arguments (see Appendix~\ref{AppendixDynamicalTime}) and is often quoted in observational studies \cite[][]{Lilly2013, Tacconi2020} as indicative of the evolution of a marginally stable rotating disk in which molecular cooling, gas fragmentation and star formation should take place.  
In Fig.~\ref{fig:tdepl_multimol_data}, H$_2$ global estimates agree well with observational results from alternative galaxy-based approaches, that suggest a best-fitting trend of 
$ t_{\rm depl, H_2}(z) = (1.6 \pm 0.5)(1+z)^{-(0.98 \pm 0.12)} \,\rm Gyr $
within 2$\sigma$ error bars (very close to $ t_{\rm dyn} $) and data dispersion of almost 1 dex
\cite[see][who used existing literature data and ALMA archive molecular-mass detections up to $z \simeq 5.3$]{Tacconi2020}.
In line with studies of cosmic cold gas showing no explicit dependence on galaxy mass, size or environment for galaxy integrated depletion times \cite[][]{Darvish2018, Tadaki2019} and notwithstanding the operational costs to determine molecular masses, H$_2$ depletion times can track cosmic gas evolution if the star formation history is well known
\cite[][]{Tacconi2020}.
Despite the relatively low scatter at $z\simeq 2$-4 (between 0.1 and 1~Gyr), lower-$z$ determinations have a larger spread.
Using available observational data, at $z\simeq 0$-2 we find $ t_{\rm depl, H_2} $ comprised between $ 10 \,\rm Gyr$ (ALMACAL-abs upper limits) and $ 0.01\,\rm Gyr $ (ASPECS and Herschel PEP lower limits).
The latest UKIDSS data at $z\simeq 2$ suggest values around 0.1-0.3~Gyr.
Interestingly, expected H$_2$ depletion times are always found below the Hubble time and, often, below the dynamical time.
This is a more stringent constraint than the one derived from neutral-gas depletion times and it supports the idea that cosmic structures form much beyond redshift $z \simeq 6$.
Thus, future observational efforts in the (sub-)mm regimes leading to larger samples of high-$z$ star forming galaxies, similar to those recently detected at $z\simeq 6$ \cite[][]{Gruppioni2020, Talia2020, Zavala2021}, are required.


\section{Discussion}\label{Sect:discussion}


In the previous sections we have investigated the role of different physical and chemical processes that could affect atomic and molecular cold-gas evolution.
We have tested three different UV backgrounds (HM, P19, FG20) coupled to non-equilibrium chemistry and found that UV radiation from forming structures has a clear impact on thermodynamics, composition and time evolution of cosmic gas.
This is generally true for all the UV scenarios, although the implications of each individual model vary remarkably and appear to be extreme for large UV onset redshifts ($z \gtrsim 8$).
\\

\noindent
Inclusion of HI and H$_2$ self-shielding is relevant to address the role of UV photoionization at high (above 1 particle cm$^{-3}$) and intermediate densities (in the range between $\sim10^{-2}$ and 1 particle cm$^{-3}$).
HI shielding has an important {\it indirect} impact during and after reionization, since it significantly preserves the amount of HI, indispensable for H$_2$ formation, from UV photoionization.
Thus, self-shielding ensures that enough amounts of HI and H$_2$ are obtained to consistently match the observations.
We remind that H$_2$ self-shielding is effective mainly on dense  gas and has a small impact on IGM/CGM-like material \cite[e.g.][]{Petkova2012}, thus HI self-shielding results to play a major role in cosmic gas chemistry.
The effect of varying H$_2$ shielding prescriptions, instead, is modest (Appendix~\ref{AppendixShielding}).
As an example, the recently updated model by \cite{WG2017} results in excellent agreement with previous works by \cite{DB1996} 
as long as the length scale used to derive the local H$_2$ column density is approximated by 1/4 the Jeans length.
The effects of alternative choices are small in thin media, but become important in dense sites \cite[e.g. when estimating molecular cooling of collapsing gas or critical fluxes of impinging radiation during the formation of massive black-hole seeds;][]{Luo2020, Maio2019}.
Neglecting HI shielding corrections at high redshift ($z > 5$) leads to a large degree of H ionization and H$_2$ photodissociation, causing under-estimates of HI and H$_2$ cosmological densities and leading to a trend of the atomic and molecular density parameters in tension with data even at $ z < 5$.
Furthermore, gas self-shielding seems crucial to justify the existence of molecular gas at high redshift, as observed ubiquitously in outflows of massive galaxies at $ z \gtrsim 4 $ with loading factors of the order of unity \cite[][]{Spilker2020}.
\\

\noindent
Gas-grain processes have been implemented adopting different values for a constant grain temperature ($T_{\rm gr} = 40$, 75 and 120~K) and a $z$-dependent one, based either on the CMB scaling or on the dust emissivity index, $\beta$ \cite[][]{daCunha2021}.
Since $T_{\rm gr}$ evolution is not very sensitive to $\beta$, we find that details of its implementation do not have a large impact on the bulk of structure formation and the overall H$_2$ mass formed. They can be relevant locally, though, in metal-rich sites \cite[in agreement with expectations from high-$z$ starburst galaxies;][]{Overzier2011} and boost molecular fractions above $50\%$ (Appendix~\ref{AppendixH2gr}).
\\

\noindent
Our results show that the origin of H$_2$ in different epochs is typically linked to free electrons made available during structure formation that boost the H$^-$ channel in most cosmic environments.
Such a relevant role of the H$^-$ channel is in line with conclusions from analytic investigations that stressed the need to reduce the density of H$^-$ to suppress star formation in early mini-haloes
\cite[][]{Cen2017}, as well as with recent semi-analytical ISM studies that highlighted the important role played by H$^-$ \cite[][]{Thi2020}.
This species is effective in electron-rich, albeit not extremely dense ($ \sim 0.1$-1~$\rm cm^{-3}$) gas, as is the case for most of the cosmic medium during or after reionization.
Free electrons are captured by H atoms via radiative attachment and increase the amount of H$^-$ available for H$_2$ formation.
The H$_2^+$ channel can provide an additional contribution (of up to $\sim 1/5 $), but it is typically smaller than that from the H$^-$ channel, due to the lower collisional rates
(Appendix~\ref{AppendixH2channels}).
Reactions involving H$^-$ and H$_2^+$ catalysts to form molecular hydrogen ($\rm H^- + H_2^+ \rightarrow H_2 + H $)
are less likely than H$_2^+$ or H$^-$ impacts with H and the resulting H$_2$ enhancement is not substantial, given also their relatively small abundances.
Three-body processes can provide some H$_2$ mass content.
Their effects are subdominant for $\rm \Omega_{\rm neutral} $ and $\Omega_{\rm H_2}$ estimates, but help reach high H$_2$ fractions of roughly $50\%$ at $z\sim 4$-6 in metal-poor sites (i.e. values similar to those reached by H$_2$ grain catalysis in metal-rich sites).\\

\noindent
These processes explain well the origin of the large supply of molecular mass within galaxies at $z\gtrsim 2$ 
\cite[][]{Garratt2021}.
In fact, large observationally inferred H$_2$ fractions at early times \cite[][]{DZ2020} can be explained either via standard H$_2$ formation channels in  metal-poor/pristine gas or via H$_2$ grain catalysis in metal-rich sites.
We warn the reader that, from a purely statistical point of view, large H$_2$ values, although detected, might not necessarily be typical for the whole galaxy population at those epochs. Indeed, there is a clear observational bias because of which galaxies with large H$_2$ fractions are easier to observe via their strong CO emission.
Previous results from sub-grid models for isolated galaxies at $z\simeq 2 $ focusing on H$_2$ formation on dust grains only \cite[][]{Tomassetti2015}
confirm the leading role of H$_2$ grain catalysis in enriched environments.
The possibility to form H$_2$ molecules via the two mentioned chemistry paths is also consistent with the lack of observed environmental (metallicity) dependence up to $z\sim 3.5$ in the assembly of H$_2$ mass
\cite[e.g.][]{Darvish2018, Tadaki2019}.
This hides the fundamental physical difference between the nearby Universe, where H$_2$ formation is led by grain catalysis in polluted media \cite[in agreement with results of local molecular clouds,][]{Wakelam2017},
and the distant one, where H$_2$ production is mainly led by the H$^-$ channel in shielded low-$Z$/pristine gas.
We warn that, in principle, H$_2$ production could be enhanced by grain processes at any $z$, as long as interested sites have local metallicities of $Z \gtrsim 0.1 \,\rm Z_\odot $.
This conclusion is consistent with detailed ISM analyses by e.g.
\cite{Gong2020} (who consider $Z$ within 0.5-2 $Z_\odot$)
and \cite{Hu2021} (who explore $Z$ within 0.1-3 $Z_\odot$).
\\

\noindent
We have improved upon previous semi-analytical studies by \cite{Jasche2007} about the possible heating due to cosmic rays at early epochs (Appendix \ref{AppendixZeta}).
Our constant cosmic-ray ionisation rate per H atoms, 
$ \zeta = 10^{-17} \, \rm s^{-1}$,
is consistent with traditional values, with the latest determinations by 
\cite{Thi2020} 
and with the rate assumed in
\cite{NarayanKrumholz2017}, who correct by a factor of 0.1 the suggested $\zeta \simeq 10^{-16} \,\rm s^{-1} $ 
\cite[][]{IndrioloMcCall2012}.
We additionally consider the calculations by \cite{Padovani2018} for a density-dependent $\zeta$ and adopt it to assess the implications on star forming gas.
We find that typical values for the cosmic-ray ionization rate induce heating in addition to the one from structure formation shocks and cause a deficit of $\Omega_{\rm H_2}$ at early times.
In the realistic scenario of cosmic-ray heating linked to star forming gas parcels, the effects observed at high $z$ are not very prominent and the resulting  $\Omega_{\rm H_2}$ and $\Omega_{\rm neutral}$ (or $\Omega_{\rm HI}$) are close to those without cosmic-ray heating.
At lower $z$, the formation of molecular gas is slightly suppressed and $\Omega_{\rm H_2}$ is a bit smaller than in the reference scenario without cosmic-ray heating.
Overall, the effects are of the order of ten per cent.
These results are consistent with the small (up to a factor of 2) star formation suppression found in latest studies of isolated disk galaxies \cite[][]{Semenov2021} and hint at the limited role of cosmic-ray heating in high-$z$ structures.
\\
 
\noindent 
The additional processes mentioned above (metallicity evolution, metal-dependent H$_2$ grain catalysis, photoelectric effect, cosmic-ray heating, etc.) may have an impact on the local environment, depending on the chemical and thermal state of hosting gas.
H$_2$ grain catalysis and photoelectric heating can boost $\Omega_{\rm H_2}$ up to a factor of 3 at $z\simeq 6$, still within observational constraints \cite[][]{Riechers2020coldz},
while large metallicities enhance H$_2$ grain catalysis up to 
$ \Omega_{\rm H_2} \simeq 2$-$3 \times 10^{-4}$ 
for $Z \sim \, Z_\odot$.
We note that we have scaled H$_2$ grain reaction and cooling/heating rates linearly with $Z$, but dust-to-gas ratios could be steeper in low-$Z$ environments (like in blue compact dwarves, which can be thought as nearby probes of the physics of distant primeval sources).
In such cases the suggested scaling could be 
$\propto Z^{1.5}$ \cite[][]{Galametz2011}
or even
$\propto Z^{2}$ \cite[][]{Herrera-Camus2012}.
This would lower the contribution of gas-grain cooling/heating at low $Z$, imply metallicities closer to solar to get H$_2$ grain catalysis effective and indicate a much lesser role of heavy elements at early times.
Non-equilibrium results about the cosmic-ray heating rely on the ionization rate and yield derived from realistic H, He and H$_2$ gas mixtures.
Although improved with respect to previous works, those quantities are still obtained by neglecting heavy elements, thus analyses of the energy transfer in more general conditions are still required.
In our study we varied it by up to one order of magnitude and found small effects for HI and H$_2$ mass densities in star forming regions at $z < 6$.
Furthermore, we tested the implications of different IMFs and wind assumptions on mass density parameters, without finding significant changes (Appendix~\ref{AppendixWindsIMF}).
\\

\noindent
The actual role of some physical processes, like photoelectric effect, cosmic rays or dust cycle in galaxies during the epoch of reionisation, is still unclear and more data need to be collected \cite[][]{Burgarella2020}.
Despite these uncertainties, gas self-shielding, that preserves HI and H$_2$ abundances, is likely the principal reason for cool molecular gas to extend further out than dust in observations of star forming galaxies.
Indeed, we checked that, without self-shielding, lower H$_2$ values are obtained even with the inclusion of H$_2$ grain catalysis.
This is in line with recent interpretations of ALMA observations \cite[]{Rivera2018} and should hold {\it a fortiori} at high redshift, when metals and dust grains are expected to be under-abundant.
In this respect observational analyses by \cite{Rivera2018} support our conclusions.

\noindent 
Earlier cosmological simulations have highlighted the difficulty in reproducing the behaviour of chemical density parameters.
In particular, the observed {\it increasing} trend of $\rm \Omega_{\rm HI}$ and $\rm \Omega_{\rm neutral}$ has been challenging to model.
Numerical efforts by \cite{Nagamine2004} using similar star formation and feedback models to \cite{SpringelHernquist2003} in a dozen Gadget simulations highlighted the relevant role of wind feedback and obtained $\rm \Omega_{\rm HI}$ results consistent with observations at $ z = 2 $-5 with a "weak" wind speed of 242~km/s. This value is in line with with our reference wind speed of 350~km/s.
The consistency in the general trend is encouraging and corroborates the results presented in this manuscript.
\cite{Nagamine2004}  could not study $ \Omega_{\rm H_2}$ evolution because of the lack of observational data at the time.
\citet[][]{Dave2013} proposed a treatment for HI and H$_2$ gas content based on H ionization balance to account for HI and on ISM pressure equilibrium to account for H$_2$\footnote{
They compute H$_2$ fractions in star forming particles only. Gas that is not star forming is assumed to have zero molecular content.}.
These approximations were required because their star formation threshold was $0.1 \, \rm cm^{-3}$ (too low to resolve runaway molecular cooling in dense cold gas) and HI/H$_2$ self-shielding effects were not included.
Gas properties were corrected for HI shielding in post-processing, but they found an HI mass density parameter decreasing with increasing redshift, with 
$\rm \Omega_{\rm HI} < 10^{-3} $ at $z < 5$ and 
$\rm \Omega_{\rm HI} \simeq$ a few $10^{-4}$ at $z\simeq 5$.
These predictions led the authors to question the ability of equilibrium-based models to capture the fundamental processes shaping the HI and H$_2$ behaviour, as remarked also in \cite{Dave2020}.
Similarly, low HI mass densities and decreasing $\rm \Omega_{\rm HI} $ redshift evolution have been obtained by other independent groups based on semi-analytical models run on top of N-body simulations
\cite[e.g.][]{Spinelli2020}.
The deviations of H$_2$ abundances from equilibrium in cosmic environments highlight the need to follow a time-dependent non-equilibrium approach to obtain reliable molecular fractions
\cite[as stressed lately by][and as done throughout this work]{Hu2017, Hu2021}.
Other popular hydrodynamic simulations 
\cite[such as Owls or Eagle;][and references therein]{Joop2015}
adopt UV rates similar to the ones used here (HM), but employ low-density thresholds for star formation and do not partition hydrogen into its ionised, neutral and molecular components \cite[as also done in the NIHAO project;][]{Obreja2019}.
So, post-processing assumptions on how to distribute these species are needed (\citealt{BR2006} or \citealt{GK2011}).
Post-processing modelling of hydrogen phases for the Eagle simulations \cite[][]{Rahmati2015} leads to a roughly flat HI curve at $z \sim 2-$6, while Illustris-TNG results in 100~Mpc/$h$ side box feature a non-monotonic behaviour attributed to lack of resolution
\cite[][]{Villaescusa2018}.
Being usually based on local calibrations, this approach is effective at studying the nearby Universe or enriched systems, but might present issues at higher redshift or in low-metallicity environments, such as the whole epoch of reionization and the first few cosmological Gyrs.
In practice, since they are not included in hydro and chemistry run-time calculations, H$_2$ molecules cannot play any role in gas cooling.
Probably for this reason, semi-analytical computations by 
\citet[and references therein]{Lagos2018}
performed on the outputs of numerical simulations to estimate HI and H$_2$ fractions with post-processing scaling relations, find HI densities which are not fully consistent with observations.
Their HI (neutral) mass density parameter decreases from a few $10^{-4} $ (at $z\simeq 1$) down to $10^{-5} $ (at $z \simeq 3.5 $) and also their values never reach $\Omega_{\rm HI} \sim 10^{-3}$
\cite[][]{Rhee2018}.
Despite that, \cite{Lagos2018} can roughly reproduce $\Omega_{\rm H_2}$ values at $z \lesssim 3$.
The various Illustris simulations \cite[e.g.][and references therein]{Springel2018} address HI and H$_2$ properties via the empirical model of \cite{GnedinDraine2014, GnedinDraine2016Erratum}, but do not implement non-equilibrium atomic and molecular chemistry.
When post-processed with semi-analitical models \cite[][]{Somerville2015} results predict H$_2$ masses that are too low at $z > 1 $ and cosmic densities ($\Omega_{\rm H_2}$) in tension with ASPECS data, independently from the adopted H$_2$ partition recipe, as their simulated gas (for which the star formation density threshold is $\sim 0.1\,\rm cm^{-3}$) is not dense enough to become molecular
\cite[][]{Popping2019}. 
The smaller $\Omega_{\rm H_2}$ values found by \cite{Popping2019} might also be linked to the chosen UV background by \cite{FG2009}, that (similarly to \citealt[FG20]{FG2020}) assumes an earlier onset and hence causes a longer photoheating injection with respect to the HM model.
Although this latter is adopted in \cite{Lagos2018}, their latest semi-analytical estimates \cite[][]{Lagos2020} still find $\rho_{\rm H_2} $ values systematically below the observational constraints.
Other recent studies (such as 
\citealt{Gjergo2018, Gjergo2020}, 
\citealt{Vogelsberger2019}, 
\citealt{Graziani2020},
\citealt{Granato2021}) 
have performed dust mass modelling from gas mass metallicities within three-dimensional numerical simulations, however their implementations are not coupled to non-equilibrium H$_2$ chemistry.
Finally, recent H$_2$ modelling by \cite{Schaebe2020} that includes H$_2$ formation via dust grain catalysis only in a 12~Mpc/$h$ side box has suggested the existence of undetected molecular amounts at high redshift, since their predicted H$_2$ mass densities at $z\simeq 4-$6 were higher than current observational determinations (as in Fig.~\ref{fig:grT}).
\\

\noindent
We note that, as in any such work, numerical accuracy is important, but the different star formation schemes and feedback implementations adopted often have a larger impact.
Feedback efficiency and star formation timescales modulate the whole evolution of H$_2$ mass.
Changing the shape of the IMF from Salpeter to Chabrier alters slightly star formation related quantities 
\cite[in agreement with e.g.][]{Cassara2013}, 
while varying wind velocities 
\cite[][]{Hassan2021} 
or SN efficiency has an impact of up to a dex when these parameters are varied by a factor of 5-10, independently from other model assumptions \cite[][]{Steinwandel2020}.
For HI the impact is less dramatic and results vary by less than half dex.
Chemistry rates and yields are still affected by factor-of-two uncertainties 
\cite[for examples compare H$_2$ grain catalysis rates in][]{HM1979, TH1985, Cazaux2004, Tomassetti2015} 
and this influences resulting abundance ratios 
\cite[][]{MaioTescari2015, Ma2015, Ma2017dla, ValentiniM2019}.\\

\noindent
From an observational point of view, HI mass densities are relatively well established up to $z\lesssim 6$ with errors below a factor of 2.
Moreover, exploiting [C~II] emission in high-$z$ enriched environments \cite[as lately done by][]{Heintz2021} would possibly open the door to observing $ \Omega_{\rm neutral }$  at $z > 6$, where our results predict different scenarios depending on the chosen UV background.
On the other hand, estimates of the reservoir of H$_2$ masses in star forming environments from [C~II] or CO observational data are challenging and usually adopted conversion factors bear uncertainties and biases
\cite[][]{Madden2020, Gong2020, Breysse2021}.
The [C~II]~158 micron line is a workhorse for (sub-)mm observations and is resolved on kpc-scale by ALMA 
\cite[][]{Rybak2019}.
Since extended [C~II] halos are found more and more commonly in high-$z$ galaxies 
\cite[early star forming sources of $10^{11}$-$10^{12}$~M$_\odot$ fit with classical `normal' star forming galaxies;][]{Ginolfi2019, Hodge2020}
and since [C~II] emissions tracing a combination of molecular, ionised and atomic gas are still very little explored, future assessments of H$_2$ via [C~II] could improve significantly the current picture for $\Omega_{\rm H_2}$.
\\

\noindent
Throughout this work we have assumed a $\rm\Lambda$CDM cosmology.
We briefly point out that in some particular cases \cite[such as for warm dark matter,][]{MaioViel2015} the background cosmological model influences the baryonic-structure evolution, mostly at higher redshift. This plays a role during the beginning of the onset of star formation, but in the epochs of interest here baryon evolution tends to dominate and alleviate any discrepancy due either to alternative cosmologies \cite[][]{Maio2006} and matter non-Gaussianities
\cite[][]{MaioIannuzzi2011, Maio2011cqg, Maio2012, MaioKhochfar2012} or to a different dark-matter nature \cite[][]{MaioViel2015}.
Finally, the baryon census ($\Omega_{0, \rm{b} }$) is crucial for assessing atomic and molecular gas evolution.
Alternative estimates of $\Omega_{0, \rm{b} }$ have been proposed \cite[using e.g. FRBs,][]{Macquart2020} and, although uncertainties still persist, they are consistent with the cosmic microwave background and the Big Bang nucleosynthesis.

\section{Conclusions} \label{Sect:conclusions}

We have quantified the evolution of cold atomic and molecular cosmic gas, from the epoch of reionization down to $z\sim 2$.
We have performed cosmological N-body hydrodynamic chemistry simulations to model the origin and evolution of cold-gas density parameters and timescales.
Our implementation follows time-dependent non-equilibrium atomic and molecular abundance calculations and considers the impact of different UV backgrounds, as well as a number of physical processes that could affect the build-up of H$_2$ and neutral-gas mass.
Star formation in dense HI and H$_2$ self-shielded regions, stellar feedback and production of heavy-elements from stars with different masses and metallicities are also accounted for.
We constrain the effects of the major processes involved in cold-gas evolution and interpret state-of-the-art observations in the (sub-)mm and IR ranges.
This is among the first studies addressing cosmic HI and H$_2$ evolution at these epochs by three-dimensional hydrodynamic simulations and including detailed non-equilibrium chemistry networks.
In summary, we find that:
\begin{itemize}
\item[--]
the evolution of cold atomic and molecular gas obtained with basic time-dependent non-equilibrium chemical reactions in dense gas is broadly consistent with the latest HI and H$_2$ observational data once HI and H$_2$ self-shielding from UV radiation is taken into account within the chemical network;
\item[--]
the resulting mass of neutral gas  slowly increases with redshift up to 
$\Omega_{\rm neutral} \sim 2\times 10^{-3} $ at $z\simeq 6$, while H$_2$ features a plateau at $\Omega_{\rm H_2} \sim 10^{-4}$ at $ z \gtrsim 2$;
\item[--]
details of the UV background model adopted (HM, P19 and FG20) have a modest impact on $\Omega_{\rm neutral}$ at $ z\lesssim 6$, while they are relevant for $\Omega_{\rm H_2}$: the onset epoch of the UV background in particular seems to be crucial, suggesting that early ($ z \gtrsim 8$) onsets could be in tension with H$_2$ data and requiring either lower UV rates or larger self-shielding at those times;
\item[--]
resolving dense shielded gas and adopting a high star formation density threshold is necessary to properly estimate the amount of cold gas and to obtain results in line with HI and H$_2$ observational data;
\item[--]
our non-equilibrium H$_2$-based star formation model leads to results that are in slightly better agreement with available HI and H$_2$ determinations;
\item[--]
H$_2$ evolution is globally driven by the H$^-$ channel, with a relevant contribution from H$_2$ grain catalysis at high metallicities and on local scales, while other formation channels are negligible: 
this explains the observationally inferred lack of environmental dependence in H$_2$ mass build-up at $z \lesssim 3.5$;
\item[--]
cosmic-ray heating in star forming regions, quantified in accordance to different models for the cosmic-ray ionization rate, affects the H$_2$ abundance at ten-per-cent level, globally;
\item[--]
wind feedback impacts results mostly at lower redshift, while SN explosions affect early star forming regimes, with effects from IMF variations remaining usually small at all times;
\item[--]
recently detected molecular fractions of $\sim 50\%$ at $z\simeq 4$-6 can be explained by essential non-equilibrium H$_2$ formation channels in metal-poor media or H$_2$ grain catalysis in metal-rich gas: in this latter case, a precise determination of grain temperature has little relevance and is generally subdominant to metallicity effects; 
\item[--]
despite the insurgence of UV radiation, H$_2$ depletion times reach values lower than the Hubble time, even comparable to the dynamical time, within the first half Gyr, meaning that cosmic gas is able to collapse already at such early epochs.
\end{itemize}
\noindent
The ability to form significant amounts of H$_2$ and to reach short depletion times during the epoch of reionization means that future discoveries of molecular-rich star forming galaxies at early times will be possible in the next years.
New data from upcoming international facilities, such as 
JWST \cite[][]{Gardner2009},
ELT \cite[][]{Puech2010},
SKA \cite[][]{Koopmans2015}, 
Roman/WFIRST \cite[][]{Whalen2013} or
ATHENA \cite[][]{Nandra2013}
will be decisive to shed light on the still unanswered questions about the origin of chemical species in the cosmic gas and the whole baryon cycle.



\begin{acknowledgements}
We are grateful to the anonymous referee for his/her careful reading and suggestions that improved the original manuscript.
We are thankful to F.~Haardt, S.~Borgani, L.~Tornatore and R.~Szakacs for their constructive comments.
U.M. is indebted to L.~B{\"o}ss and U.~Steinwandel for insightful conversations and to M.~Padovani for kindly sharing the full table of his ionisation rate calculations used throughout this work.
We thank A.~Hamanowicz for sharing with us preliminary ALMACAL-CO results.
We also thank the Max Planck Computation and Data Facility (MPCDF) of the Max Planck Society, where simulation runs and analyses have been performed.
We acknowledge the NASA Astrophysics Data System and the JSTOR archive for their bibliographic tools.

\end{acknowledgements}


\bibliographystyle{aa1}
\bibliography{bibl}


\appendix

\section{Data samples}  \label{AppendixData}

Data samples for H$_2$ mass density values, $ \rho_{\rm H_2} $ (and corresponding $ \Omega_{\rm H_2} $), are summarised in Tab.~\ref{tab:dataOmegaH2}, where redshift range, ${\rm H_2} $ measurements and publication references are quoted.
Many of them are present in \cite{PerouxHowk2020}.

\begin{table*}
\small
\centering
\caption{
Observational determinations of molecular (H$_2$) gas considered in this work.
}
\label{tab:dataOmegaH2}
\vspace{-0.2cm}
\begin{tabular}{c c c c c c c c c}
$ z_- $	&	$ z_+ $	&	Log $\rho_{{\rm H_2},-} $	& 	Log $\rho_{{\rm H_2}, +}$ &	Log $ \Omega_{{\rm H_2}, -} $	& 	Log $ \Omega_{{\rm H_2}, +} $ &	Reference &	Survey &	 \\
\hline
0.0		& 	0.5		&	5.903	&	6.940	&	-5.230 	& 	-4.194	& 	\cite{Hamanowicz2021}$^{\star}$  &	 ALMACAL-CO & \\
0.5		& 	1.0 	&	6.991	&	7.510	&	-4.142 	& -3.623	&  "	&	"	& \\
1.0		& 	1.5		&	6.832	&	7.420	&	-4.301 	& -3.714	& "	&	"	& \\
0.0 	& 	0.5		&	6.362	&	7.228	&	 -4.772	& 	-3.906	& "	&	"	& \\
0.0		&	0.5		&	5.08	&	6.57	&	-6.05	&	-4.56 	&	\cite{Garratt2021}  & 	UKIDSS-UDS & \\
0.5		&	1.0		&	7.10	&	7.33	&	-4.03	&	-3.80 	&  " & "	& \\
1.0		&	1.5		&	7.28	&	7.44	&	-3.85	&	-3.69	&  " & "	& \\
1.5		&	2.0		&	7.47	&	7.60	&	-3.66	&	-3.53 	&  " & "	& \\
2.0		&	2.5		&	7.21	&	7.40	&	-3.92	&	-3.73 	&  " & "	& \\
-		&	0 		&	6.581	&	6.992	&	-4.553	&	-4.142 	&	\cite{Fletcher2021}  & 	xCOLDGASS & \\
0.003 	&	0.369	&	5.17	&	6.36	&	-5.96	&	-4.77 	&	\cite{Decarli2020} 		& ASPECS	& \\
0.271	&	0.631	&	6.78	&	7.33	&	-4.35	&	-3.80 	& "	& 	"	& \\
0.695	&	1.174	&	7.44	&	7.87	&	-3.69	&	-3.26	& "	& 	"	& \\
1.006	&	1.738	&	7.61	&	7.87	&	-3.52 	&	-3.26	& "	& 	"	& \\
2.008	&	3.107	&	7.27	&	7.65	&	-3.86	&	-3.48	& "	& 	"	& \\
3.011	&	4.475	&	7.20	&	7.52	&	-3.93	&	-3.61	& "	& 	"	& \\
2.0		&	3.0		&	6.96	&	7.80	&	-4.17 	& 	-3.33	&	\cite{Riechers2020vlaspec} 	& VLASPEC	& \\
1.95	& 	2.85	&	7.041	&	7.748	&	-4.093	&	-3.386 	&	\cite{Riechers2020coldz} &	COLDz & \\
4.90	& 	6.70	&	6.146	&	7.041	&	-4.988	&	-4.093	& "	& 	"	& \\
0.4799	& 	1.4799	&	7.47	&	7.72	&	-3.66	&	-3.41	&	\cite{Lenkic2020}	  &	PHIBBS-2 & \\
1.0056	& 	2.0056	&	7.56	&	7.78	&	-3.57	&	-3.35	& "	& 	"	& \\
2.0084	& 	3.0084	&	7.42	&	7.67	&	-3.71	&	-3.46	& "	& 	"	& \\
3.1404	& 	4.1404	&	6.91	&	7.34	&	-4.22	&	-3.79	& "	& 	"	& \\
4.2451	& 	5.2451	&	6.58	&	7.22	&	-4.55	&	-3.91	& "	& 	"	& \\
0.003	&	0.369	&	-		&	<8.26	&	- 		& <-2.87	& 	\cite{Klitsch2019}	&	ALMACAL-abs&  \\
0.2713	&	0.6306  &	-		&	<8.32	&	- 		& 	<-2.81	& "	& "	&  \\
0.6950	&	1.1744  &	-		&	<8.39	&	- 		& 	<-2.74	& "	& " &  \\
1.006	&	1.738  &	-		&	<8.21	&	- 		& 	<-2.92	& "	& "	&  \\
2.0	&	3.0	&	6.00	&	-	&	-5.14 	& -	& \cite{Berta2013}	&	Herschel PEP & \\ 
\hline
\end{tabular}
\begin{flushleft}
From left to right, columns refer to 
lower and upper redshift range limits,
H$_2$ mass density $\rho_{\rm H_2} \rm [M_\odot/Mpc^3]$ with corresponding
$\Omega_{\rm H_2}$ range limits (1$\sigma$),
publication reference from which H$_2$ data are taken or converted and
observational survey.
\\
$^\star$ Submitted. Private communication.

\end{flushleft}
\vspace{-0.5cm}
\end{table*}

\section{HI and H$_2$ self-shielding}  \label{AppendixShielding}

\begin{figure}
\includegraphics[width=0.4\textwidth]{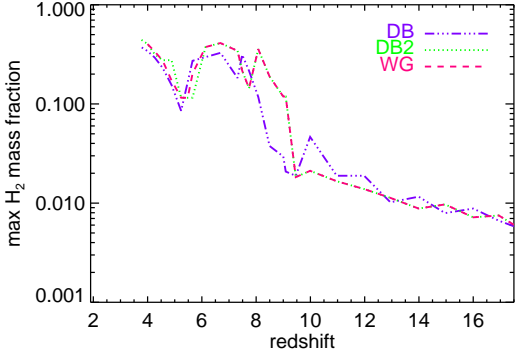}
\caption[]{\small
Redshift evolution of the maximum H$_2$ abundance reached in simulations with different H$_2$ self-shielding formulations:
eq.~36 (violet dot-dot-dot-dashed line; HM-HISSmed-3b in Tab.~\ref{tab:properties}) and 
eq.~37 (green dotted line; HM-HISSmed-3b-DB2)
in \cite{DB1996} and the formulation by \cite{WG2017} (magenta dashed line; HM-HISSmed-3b-WG).
}
\label{fig:H2shielding}
\end{figure}

We write the HI self-shielding, $f_{\rm shield, HI} $, as a function of H number density, $n_{\rm H}$, as:
\begin{equation}
\label{fshield_fit}
f_{\rm shield, HI} = 
0.985 
\left[ 1 + \left( \frac{n_{\rm H}}{n_0} \right)^{1.9} \right]^{-2} +
0.015 
\left( 1 + \frac{n_{\rm H}}{n_0} \right)^{-0.72},
\end{equation}
where $n_0 = 0.006 \,\rm cm^{-3}$ is a parameter.
Asymptotically $f_{\rm shield, HI} $ converges to 1 for $n_{\rm H} \ll n_0 $ and to 0 for $n_{\rm H} \gg n_0 $.
Besides eq.~\ref{fshield_fit}, we have checked that the alternative use of the tables by \cite{Rahmati2013} does not change substantially our findings, although truncation at $z = 5$ causes unstable behaviours in abundance calculations around that particular redshift.
\\
For H$_2$ shielding we refer to \cite{DB1996} (their eq.~36 in most cases and their eq.~37 in selected runs).
Alternative formulations for H$_2$ shielding, such as the revisitations by \cite{Sternberg2014} or \cite{WG2017} of \cite{DB1996}'s expressions, have a similar impact on the $ \Omega_{\rm H_2} $ mass density parameter.
A detailed comparison of different H$_2$ shielding models is also done in \cite{Luo2020}.
Maximum deviations from the reference run, reached locally at different epochs, can be estimated by looking at the maximum H$_2$ mass fractions recorded at different $z$.
These are more sensitive than $\Omega_{\rm H_2} $ and are plotted in Fig.~\ref{fig:H2shielding} for simulations including H$^-$, H$_2^+$ and three-body formation channels, with H$_2$  self-shielding parameterizations by \cite{DB1996} and the more recent \cite{WG2017} one.
Although we observe up to an order of magnitude differences at selected redshifts, overall the maximum H$_2$ mass fraction reached is not substantially affected by the prescription used to model the H$_2$ self-shielding.

\section{H$_2$ grain catalysis} \label{AppendixH2gr}

\begin{figure}
\includegraphics[width=0.4\textwidth]{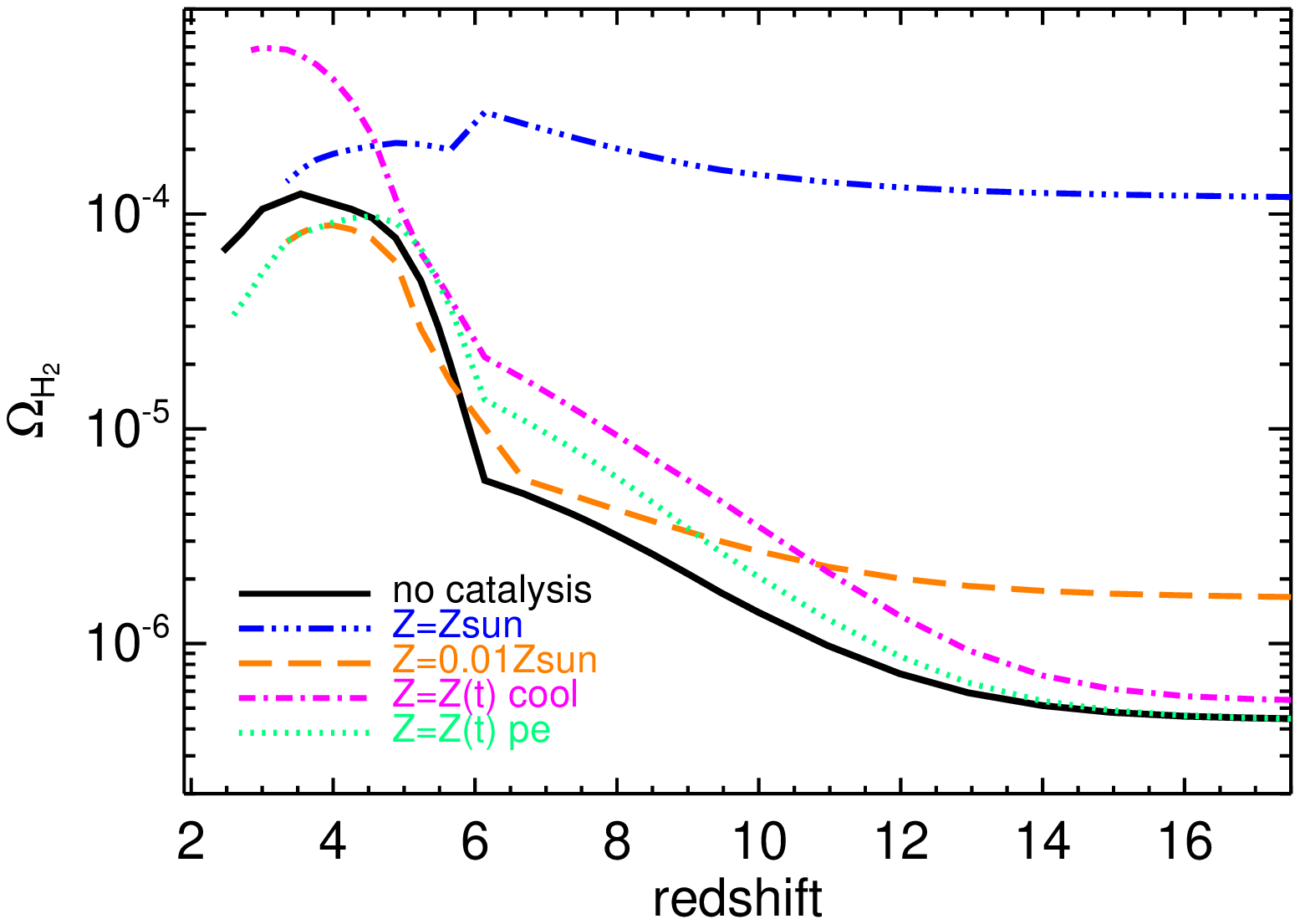}\\
\includegraphics[width=0.4\textwidth]{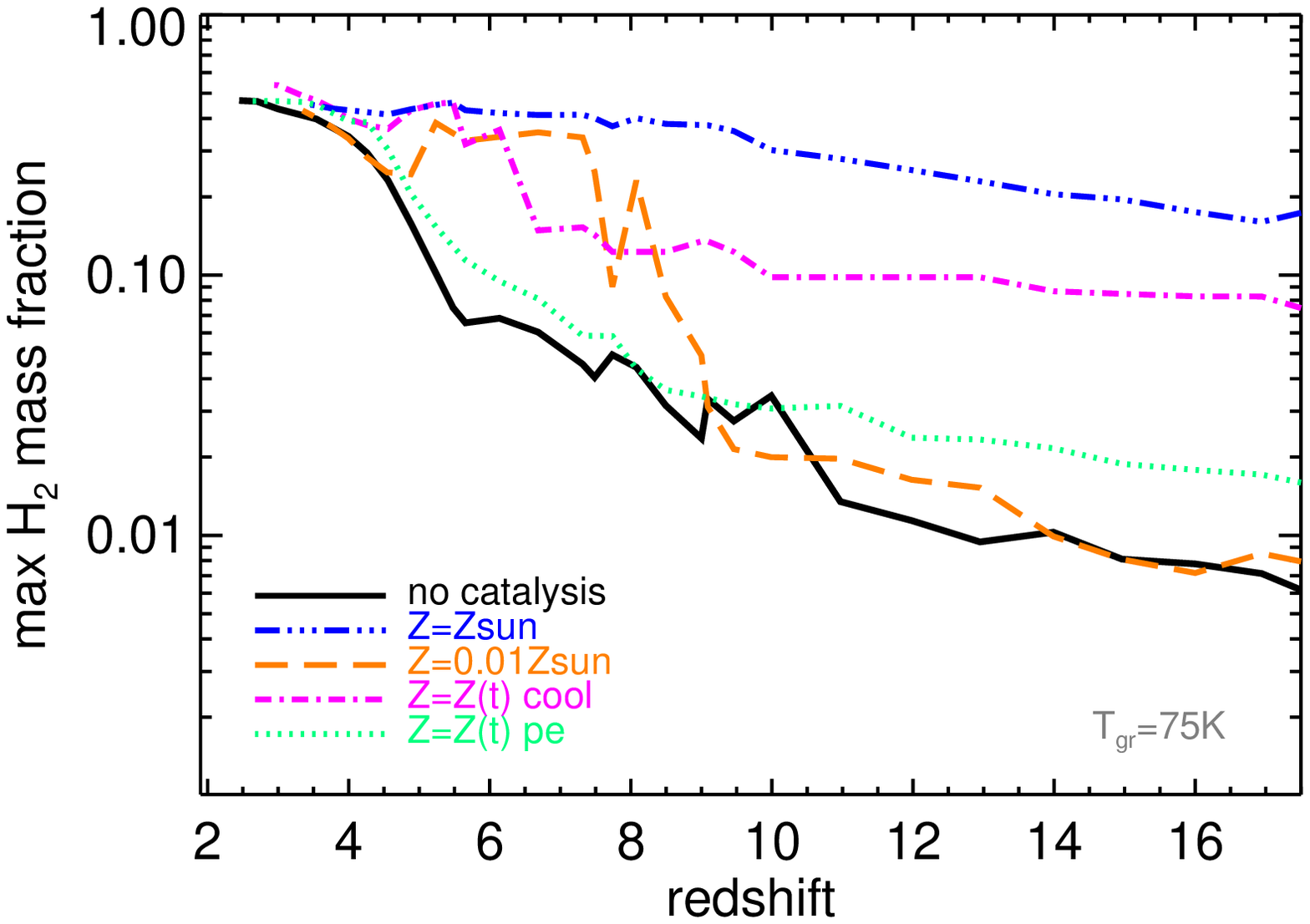}
\caption[]{\small
$ \Omega_{\rm H_2}$ evolution (upper panel) and maximum H$_2$ mass fractions (lower panel) 
reached in the no-catalysis simulation (black solid lines; HM-HISSmed) 
are compared to runs including H$_2$ grain catalysis at $ T_{\rm gr} = 75 \,\rm K $.
Gas $Z$ is either forced to constant values of 
$ Z = Z_\odot $ (blue dot-dot-dot-dashed lines; HM-HISS-cat-75-Zsun) and 
$ 0.01\,Z_\odot $ (orange dashed lines; HM-HISS-cat-75-0.01Zsun),
or consistently computed from temporal evolution of SN~II, AGB and SN~Ia metal spreading and cooling from both resonant and fine-structure lines (magenta dot-dashed lines; HM-HISS-cat-75-Zt).
Photoelectric heating on dust grains is tested, as well (green dotted lines; HM-HISSmed-cat-75-pe).
}
\label{fig:grT}
\end{figure}

The rate of H$_2$ formation due to dust grain catalysis at gas temperature $T$ and grain temperature $T_{\rm gr}$ is:
\begin{equation}
\label{eq:kH2gra}
k_{\rm H_2, gr} = 
   3.5\times 10^{-17} 
\, S_{\rm H} 
\sqrt{ \frac{T}{100~{\rm K}}}
\, {\rm cm^3/s},
\end{equation}
where the sticking coefficient $S_{\rm H}$ is given by  \cite[][]{HM1979, Cazaux2004}:
\begin{equation}
\frac{1}{S_{\rm H}} = 
    1 + 
    0.4 \sqrt{ \frac{T + T_{\rm gr}}{100~{\rm K}} } + 
    0.2 \left(\frac{T}{100~{\rm K}}\right) + 
    0.08 \left(\frac{T}{100~{\rm K}} \right)^2.
\end{equation}
In the absence of precise determinations at different epochs and $Z$, 
$ k_{\rm H_2, gr} $ is linearly scaled by $Z$, via 
$ k_{\rm H_2, gr} \rightarrow k_{\rm H_2, gr} \times \, Z /\rm Z_\odot $.\\
The normalization of eq.~\ref{eq:kH2gra} is similar to the one in \cite{Thi2020} and \cite{Sternberg2014}, while \cite{Omukai2005} divide eq.~\ref{eq:kH2gra} by 
 $ 1 + {\rm exp}{ [ 750(1/75 - 1/T_{\rm gr}) ] }$, 
which suppresses catalysis by a factor $ \ge 2$ at grain temperatures $T_{\rm gr} \ge 75\,\rm K $.
The energy density transfer rate, $\Lambda_{\rm gr} $, is computed accordingly.
In the canonical form, calibrated in the near Universe, 
\begin{equation}
\label{eq:LambdaH2gra}
\Lambda_{\rm gr}
  \simeq 3 \times 10^{-33}\sqrt{T}\left( T-T_{\rm gr} \right) \,n_{\rm H}^2  \, \rm erg/s/cm^3.
\end{equation}
Also in this case there are alternative formulations, as e.g. in \citealt{Omukai2005}, where the expression is multiplied by the dimensionless factor 
$ 1 - 0.8 {\rm exp}{ (\rm -75\,K/T) } $.
Similarly to the H$_2$ formation rate, the energy density transfer rate is scaled by $Z$,
$ \Lambda_{\rm gr} \rightarrow \Lambda_{\rm gr} \times \, Z /\rm Z_\odot $.
We ran simulations by employing both  eqs.~\ref{eq:kH2gra}-\ref{eq:LambdaH2gra}
and corrections by \cite{Omukai2005}, finding differences of at most few per cent for the H$_2$ fractions and median relative variations of the order of $10^{-4}$ for $\Omega_{\rm H_2}$ values.
For HI and neutral gas differences are even smaller.\\
\begin{figure}
\includegraphics[width=0.4\textwidth]{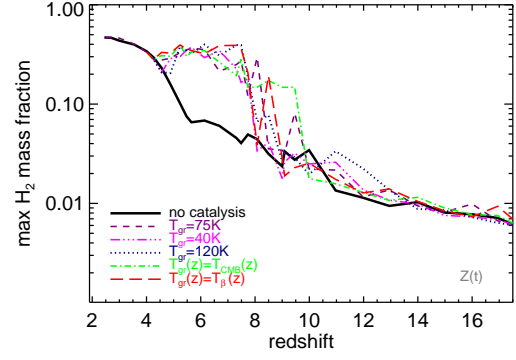}
\caption[]{\small
Redshift evolution of the maximum H$_2$ fractions reached in simulations without (solid black line; HM-HISSmed reference run) and with H$_2$ dust grain catalysis for 
$T_{\rm gr} = $ 
 75~K (purple short-dashed line; HM-HISS-cat-75-Zevol), 
 40~K (magenta dot-dot-dot-dashed line; HM-HISS-cat-40-Zevol), 
 120~K (blue dotted line; HM-HISS-cat-120-Zevol), 
$T_{\rm gr}(z) = T_{\rm CMB} (z)$ 
(green dot-dashed line; HM-HISS-cat-Tcmb-Zevol) 
and
$T_{\rm gr}(z) = T_{\beta} (z)$ with  dust emissivity index $\beta=2$
(red long dashed line; HM-HISS-cat-Tbeta-Zevol) 
are also considered.
All the runs include metal spreading from stellar evolution, as indicated in the bottom right corner by $Z(t)$.
}
\label{fig:grTfractions}
\end{figure}
In Fig.~\ref{fig:grTfractions} we explore the effect of $ T_{\rm gr} $, by showing the evolution of the H$_2$ abundance obtained in simulations with the reference no-catalysis implementation, together with runs including dust grain catalysis with 
$T_{\rm gr} = 40$, 75 and 120~K, 
as well as a $z$-dependent $T_{\rm gr}(z)$ following
either 
the CMB, $T_{\rm gr}(z) = T_{\rm 0, CMB } (1+z) $ 
with $T_{\rm 0, CMB} \simeq 2.7 \,\rm K$,
or the dust emissivity model with $\beta = 2$ \cite[][]{daCunha2021}.
We note that here three-body interactions for H$_2$ production (considered e.g. in Fig.~\ref{fig:H2shielding} and \ref{fig:H2via3H}) are switched off.
We see that grain catalysis is responsible for increasing maximum H$_2$ fractions at $z\simeq 4$-8 up to $\sim 50$\% (corresponding to roughly 67\% of H mass).
This holds independently from the  exact value of $T_{\rm gr} $, or whether it evolves with redshift.
$ \Omega_{\rm neutral} $ and  $\Omega_{\rm H_2} $ are unaffected by the exact choices of constant $T_{\rm gr}$ and are weakly affected from different scaling relations -- 
e.g. $T_{\rm CMB}$ vs. $T_{\rm \beta}$ -- 
for $T_{\rm gr}(z)$ evolution.
\\
In Fig.~\ref{fig:grT} we explore the effects of metallicity on H$_2$ grain catalysis at $ T_{\rm gr} = 75\,\rm K $.
Indeed, such H$_2$ formation process is efficient mostly in cold collapsing gas after metal pollution has started.
In addition to the reference simulation, we show results obtained from test runs in which we assume H$_2$ grain catalysis at fixed
$Z = \rm Z_\odot$ and 
$Z = 0.01 \, \rm Z_\odot$.
The largest effects are expected at $ z\gtrsim 5 $, where metallicities have a remarkable impact on $\Omega_{\rm H_2}$.
Typical metallicities in the no-catalysis case are close to zero at $z\sim 20$ and reach values around $ 0.01 \, \rm Z_\odot$ at $z\sim 10$, but they have no direct effects on H$_2$, because grain catalysis is not active.
In the law-$Z$ case ($Z=0.01\,\rm Z_\odot$) one sees the formation of an $\Omega_{\rm H2} \sim 2 \times 10^{-6} $ floor at high redshift that converges at $z<6$ to the reference trend.
The formation of such a floor is due to additional H$_2$ formed via grain catalysis in early enriched gas.
The increasingly dominant role of UV radiation at $z\lesssim 6$ provides free electrons that enhance H$_2$ formation in shielded and/or in recombining gas.
This is consistent with a leading role played by the H$^-$ channel, which becomes more important than law-$Z$ gas-grain processes at the end of reionization.
The extreme $Z=Z_\odot$ case predicts 
$ \Omega_{\rm H_2} \sim 1$-$3 \times 10^{-4}$ 
during most of the cosmological evolution.
These values are higher than those expected in the reference run, but one should keep in mind that they are probably unrealistic, as it is unlikely that most of the cosmic gas is polluted at solar metallicity by the first (fractions of) Gyr.
In any case, it is a nice confirmation that at high $Z$ grain catalysis seems to be the mostly relevant process for $\Omega_{\rm H_2}$.
The effects of UV background onset around $z\simeq 6$ (with the additional contributions from the H$^-$ channel) causes modest adjustments to the otherwise smooth $\Omega_{\rm H_2} $ evolution.
The results obtained with a constant $Z$ highlight how assuming a fixed metallicity floor at early times could drive to misleading conclusions for H$_2$ estimates (off by orders of magnitude) in the first few billion years. \\
Maximum H$_2$ abundances get clearly boosted at any redshift in the (extreme) case with constant $ Z = Z_\odot $.
In the case with $ Z = 0.01\,Z_\odot $ floor, instead, H$_2$ fractions are quite close to the results expected for the reference (no-catalysis) run at $z \gtrsim 10 $, but, consistently with Fig.~\ref{fig:grTfractions}, they show an enhancement at $z\simeq 4$-8, when H$_2$ mass fractions reach $\sim 50$\% (i.e. 67\% with respect to H mass).
At $z \gtrsim 8$, H$_2$ fractions are similar in both cases, but corresponding $\Omega_{\rm H_2}$ values (upper panel) differ as a consequence of the adopted $Z$ floor that enhances H$_2$ production by a tiny amount in {\it all} the gas parcels. The cumulative result tends to an increase of  $\Omega_{\rm H_2} $ by a factor of a few at $z\gtrsim 12$.
Results from the run including H$_2$ grain catalysis at $T_{\rm gr} = 75 \,\rm K$ and time-dependent metallicity, $Z = Z(t)$, derived from consistent metal evolution and cooling from both resonant and fine-structure lines, show a slightly larger $\Omega_{\rm H_2}$ than the no-catalysis one (due to the generally increasing global metallicities) and an intermediate behaviour of the maximum H$_2$ fractions (resulting from early local $Z$ values typically comprised within $\sim 10^{-2}$-1 $Z_\odot $; e.g. \citealt{Maio2010}).
$ \Omega_{\rm neutral} $ is little affected.

\section{photoelectric heating}   \label{AppendixPE}

The photoelectric effect in the interstellar medium is caused by the absorption of UV radiation (usually produced by star formation events) from dust grains, which then re-emit electrons and heat up the surrounding medium.
Therefore, the power density of photoelectric heating, $\Gamma_{\rm pe} $, is a function of the UV field, parameterized through the dimensionless \cite{Habing1968} parameter $G_0 \simeq 1.7$, gas temperature, $T$,  and electron number fraction, $n_{\rm e} $ 
\cite[][]{DraineSutin1987, BakesTielens1994, Weingartner2001}.
The efficiency of the process, $\varepsilon$, depends on $ G_0 \sqrt{T} / n_{\rm e} $ which includes the effects of radiation, recombination rate and electron number density:
\begin{equation}
\varepsilon(\psi) = 
\frac{0.0487}{1 + 0.004\,\psi^{0.73} } + 
    \frac{ 0.0365 ( T/10^4\,\rm K)^{0.7}  }{1 + 0.0002\,\psi }
\end{equation}
where $\psi = G_0 \sqrt{T} / n_e $ in units of $\rm K^{1/2} \, cm^3$.
The corresponding power density is 
\begin{equation}
\Gamma_{\rm pe} = C_{\rm pe} ~\varepsilon(\psi)~G_0~n_{\rm H},
\end{equation}
where
$C_{\rm pe} \simeq 10^{-24 } \,\rm erg/s $ is a normalization factor estimated in the nearby Universe and $n_{\rm H}$ is the H number density.
Because of its dependencies, a quantitative assessment of photoelectric heating needs to be done for all the variety of conditions in which the cosmic gas could possibly be found.
Qualitatively, since $\varepsilon $ depends inversely on $\psi $ for any fixed $G_0$, gas at high (low) temperatures experiences limited (relevant) photoelectric heating.
Similarly, for a given thermal state, high (low) UV fluxes lead to enhancement (quenching) of $\Gamma_{\rm pe}$.
Due to a lack of information about photoelectric heating in the distant Universe, $G_0$ is either kept constant or scaled by the SFR i.e. $ G_0 \rightarrow G_0 \times \rm SFR/SFR_\odot $, where $ \rm SFR_\odot $ is the Milky Way value.
We tried both approaches without finding large differences.
As we have coupled photoelectric heating with the chemistry network in a flexible way, alternative prescriptions could be easily tested.\\
To understand the role of this process during cosmological structure formation, in Fig.~\ref{fig:grT} we over-plot results for a run including both H$_2$ grain catalysis with $Z$ given by the actual temporal metallicity evolution and photoelectric heating (dotted lines).
The net effect on H$_2$ mass fractions and mass density parameters results from the trade-off between H$_2$ enhancement (due to dust grain catalysis) and H$_2$ dissociation (due to the additional heating mechanism), as clear from the trends in the figure.
Compared to the reference no-catalysis case, the final H$_2$ mass fractions reached (lower panel) in presence of both photoelectric heating and H$_2$ grain catalysis is not larger than a factor of 2.
Albeit modest, this enhancement interests regions that are increasingly enriched during cosmic time and progressively more abundantly distributed in cosmic space.
The cumulative effect on $\Omega_{\rm H_2} $ (upper panel) is marginal at $z \gtrsim 10$, but becomes visible later on, causing a steeper trend at $z\simeq 6$-10.
At $z\lesssim 6$, newly available free electrons from re-ionization lead, also in this case, H$_2$ formation.\\
Compared to the case including metal cooling (dot-dashed lines), photoelectric effect is subdominant.
$\Omega_{\rm neutral}$ is little changed.

\section{ Fitting formula for cosmic-ray ionization rate $\zeta(N)$ and H$_2$ mass build-up } 
\label{AppendixZeta}

\begin{figure}
\includegraphics[width=0.4\textwidth]{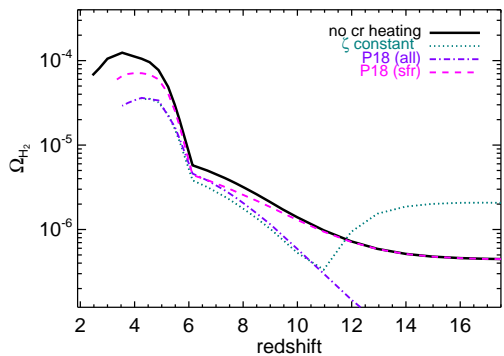}
\caption[]{\small
$\Omega_{\rm H_2}$ evolution when cosmic-ray heating included in the reference implementation (black solid line, HM-HISSmed).
The case of a constant homogeneous $\zeta = 10^{-17} \,\rm s^{-1}$ (turquoise dotted line, HM-HISSmed-cr) is displayed together with two cases in which \cite{Padovani2018}'s $\zeta$ is adopted either for all gas parcels (violet dot-dashed line, HM-HISSmed-crP18) 
or scaled by the local gas SFR (magenta dashed line, HM-HISSmed-crP18sfr).
}
\label{fig:omegaCR}
\end{figure}

A fit for the \cite{Padovani2018}'s tabulated $\mathcal{L}$ model as a function of the  gas column density, $N$, in units of $10^{19}\,\rm cm^{-2}$, $N_{19} = N / (10^{19}\,\rm cm^{-2}) $, is given by:
\begin{equation}
\label{ZetavsN}
\zeta(N_{19}) = a \,\, N_{19}^\alpha \,\,  \left( 1+ \frac{N_{19}}{b} \right)^\beta {\rm e}^{-(N_{19}/c)^\gamma},
\end{equation}
where
$ a = 3.7221566 \times 10^{-16} \,\rm s^{-1} $ is the normalisation required to reproduce the \cite{Padovani2018} $\zeta$ value at $N=10^{19}\,\rm cm^{-2}$ ($N_{19} = 1$).
The other dimensionless parameters are:
$ b = 10^2 $,
$ c = 2 \times{10^6} $,
$ \alpha = - 1/2$,
$ \beta = 1/3$ and
$ \gamma = 7/10$.
So, $\zeta$ scales as $N_{19}^{-1/2}$ at $N_{19} \ll b$, 
evolves roughly as $ N_{19}^{-1/6}$ in the range $ b \ll N_{19} \ll c$
and decays exponentially at $N_{19} \gg c$.
Since this formulation is tested on the nearby Universe, it is customal to scale $\zeta$, which is mostly related to SN events, by the gas SFR with respect to the local one, SFR$_\odot$, i.e. 
$\zeta \rightarrow \zeta \times \, \rm SFR / SFR_\odot$
 \cite[e.g.][]{NarayanKrumholz2017}.
\\
We test the effects of cosmic-ray heating on the H$_2$ evolution by employing a constant homogeneous ionization rate of $\zeta \simeq 10^{-17}\,\rm s^{-1}$ per H atom, as well as the density-dependent one by \cite{Padovani2018} 
(eq. \ref{ZetavsN}).
We also run this latter case considering cosmic-ray heating only in star forming regions, in which $\zeta$ is scaled by the gas SFR.
The resulting evolution of the H$_2$ mass density is displayed in Fig.~\ref{fig:omegaCR}, where we additionally show the case with no cosmic-ray heating for comparison.
In general, the presence of a cosmic-ray heating floor (as in e.g. the $\zeta$ constant case, dotted line) enhances the amounts of free electrons already at early times and this can increase the formation of H$_2$ at $z \gtrsim 12$ by a factor of $\sim$4.
Nevertheless, the additional heating from ongoing structure formation at later times leads to negative feedback effects and causes the subsequent decrease of $\Omega_{\rm H_2} $.
As a result, at $ z < 12 $ the system reaches a new equilibrium for H$_2$ creation and destruction processes and $\Omega_{\rm H_2} $ stays below the reference no-heating trend by almost a factor of 3 at $z\simeq 4$.
The \cite{Padovani2018} rate is stronger by up to more than one dex, thus, if applied to all particles, it creates a heating floor that easily ionizes gas and dissociates H$_2$ even at higher redshifts.
At lower redshift the resulting $\Omega_{\rm H_2} $ behaviour converges to the constant $\zeta$ case.
Applying such model to star forming particles only smooths the extreme behaviour at early times and allows us to recover an evolution of $\Omega_{\rm H_2}$ at low $z$  that is similar to, albeit slightly (less than a factor of 2) smaller than, the reference one at $ z < 6 $.
We conclude that cosmic-ray heating may affect H$_2$ abundances during the epoch of reionization and afterwards.
The impact is up to few-tens-per-cent levels for the more realistic SFR-linked case.
We note that cosmic-ray heating in the early Universe typically inhibits the H$_2^+$ channel.
Indeed, through interactions with cosmic-ray protons, H$_2^+$ is quickly converted into H$_3^+$.
This latter is an unimportant coolant that contributes no more than a few percent of the total cooling \cite[][]{Glover2009} and at large densities, where the gas tends to become fully neutral, suffers from H$^+$ removal, dissociative recombination and H-impact dissociation.
Maximum molecular fractions reached are consistent with those derived in the previous cases as well as the density parameters of HI/neutral gas for the SFR-linked case.
In the other cases $\Omega_{\rm neutral }$ values at $z<5$ converge to slightly lower values than the reference ones.
While the homogeneous presence of cosmic-ray heating is an idealized test case, we can consider as a fiducial model the one including cosmic-ray heating in star forming regions according to the \cite{Padovani2018}'s description.

\section{H$_2$ formation channels}
\label{AppendixH2channels}

\begin{figure}
\includegraphics[width=0.4\textwidth]{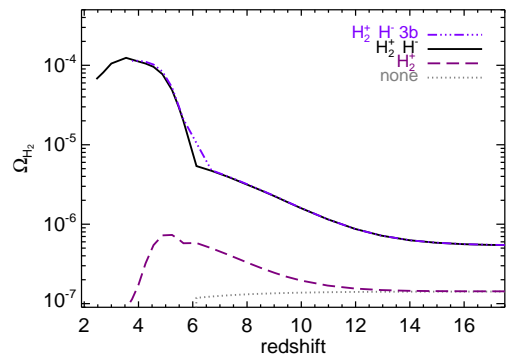}\\
\includegraphics[width=0.4\textwidth]{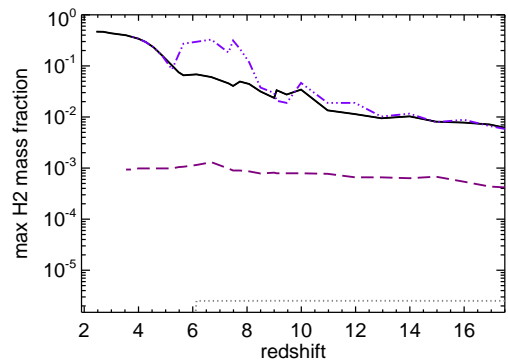}
\caption[]{\small
Redshift evolution of $ \Omega_{\rm H_2} $ (upper panel) and H$_2$ (lower) fractions obtained in the reference run 
(black solid line; HM-HISSmed in Tab.~\ref{tab:properties}), as well as in simulations including also three-body processes 
(violet triple-dot-dashed line; HM-HISSmed-3b), 
only the H$_2^+$ channel 
(purple dashed line; HM-HISSmed-H$_2^+$),
and none of the H$_2$ channels 
(grey dotted line; HM-HISSmed-none).
}
\label{fig:H2via3H}
\end{figure}

H$_2$ formation can be led by different pristine-chemistry processes.
To understand which of these is the main driver of $\Omega_{\rm H_2} $ evolution, we examine the effects of the H$_2^+$, H$^-$ and three-body channels
(for the latter we follow 
\citealt{Forrey2013}, 
which improves on previous studies by e.g.
\citealt{Palla1983},
\citealt{Orel1987},
\citealt{Abel2002} and
\citealt{Flower2007}).
In all the runs we include HI shielding (eq.~\ref{fshield_fit}) and H$_2$ shielding \cite[][]{DB1996}.
Fig.~\ref{fig:H2via3H} displays the evolution of $\Omega_{\rm H_2}$ (top panel) and of the maximum H$_2$ mass fraction (bottom panel).
Three-body interactions dominate H$_2$ formation in dense, cold regions, where HI atoms are abundant, however they do not impact significantly the global $\Omega_{\rm H_2} $ values reached by H$_2^+$ and H$^-$ channels (upper panel).
Compared to the case with only H$_2^+$ and H$^-$ channels, three-body processes locally enhanced H$_2$ formation by a factor of almost 10 at $z\lesssim 8$, raising the H$_2$ mass fractions to about 50\% by the end of reionization (lower panel).
To understand better the relative role played by the H$^-$ and H$_2^+$ channels, in Fig.~\ref{fig:H2via3H} we show results from simulations including only the H$_2^+$ channel or neither.
When no H$_2$ formation channel is considered, both $\Omega_{\rm H_2}$ and H$_2$ fractions remain practically constant at the initial-condition values.
Adding the H$_2^+$ formation channel increases both the H$_2$ density parameter and the H$_2$ fraction, but resulting values are still well below the required ones.
Reliable values are obtained by including, besides the H$_2^+$, the H$^-$ channel, which appears to be crucial for H$_2$ pristine chemistry.
This can be understood in light of the sparsity of high-density regions (this limits the role of three-body processes), the large amount of cosmic gas at moderate densities (above 0.1-$1 \,\rm cm^{-3}$) with temperatures $\lesssim 10^4\,\rm K$ and the larger reaction rates of the H$^-$ channel with respect to the H$_2^+$.

\section{Comparison of UV background effects} 
\label{AppendixOmegaUV}

\begin{figure}
\includegraphics[width=0.4\textwidth]{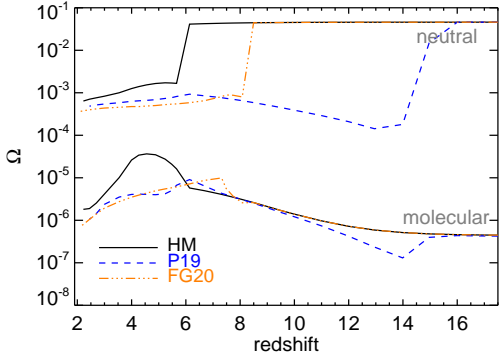}
\caption[]{\small
Redshift evolution of the neutral and molecular density parameters for the HM (solid black lines), P19 (dashed blue) and FG20 (dot-dot-dot-dashed orange) simulations, respectively.
}
\label{fig:OmegaUV}
\end{figure}
\noindent
In Fig.~\ref{fig:OmegaUV} we quantify the impact of different UV backgrounds on the amount of cold gas, showing the evolution of $\Omega_{\rm neutral}$ and $\Omega_{\rm H_2}$ for un-shielded HM, P19 and FG20 runs with the basic implementation.
For all the UV background scenarios, $\Omega_{\rm neutral}$ is higher than $\Omega_{\rm H_2}$ at all redshifts.
At high $z$, before the onset of photoheating processes, the HI content is much larger than the one of H$_2$, reflecting the primordial chemical composition of a mostly neutral cosmic gas with neutral fractions close to the baryon fraction ($ \Omega_{0,\rm{b} } \simeq  0.045 $) and molecular fractions around $10^{-6}$.
At later times, $\Omega_{\rm neutral}$ decreases and $\Omega_{\rm H_2}$ increases with a relative ratio shrinking by a few dex.
The displayed trends are sensitive to the onset of the UV background at $z\simeq 6$, 8.3, 15.1 for HM, FG20 and P19, respectively, when the transition from a neutral to an ionized phase is initiated.
Then, $\Omega_{\rm neutral}$ deviates from its initial value $\sim 0.045$, while molecular gas responds to the new thermal conditions.
The HM and FG20 scenarios produce similar results at $z \gtrsim 8$ (when neither background is turned on), but at lower $z$ the larger FG20 rates\footnote{
For example, FG20 HI and HeI rates at $z\simeq 6$ are
$\sim 10^{-12}  \,\rm s^{-1} $, 
while HM rates are below 
$10^{-15} \,\rm s^{-1} $.
At the same $z$, FG20 HI and HeI heating power is above
$10^{-24} \,\rm erg/s $, 
while the HM ones are below 
$ 10^{-26} \,\rm erg/s $.
}
are able to ionize gas and dissociate molecules more efficiently than the HM.
A peculiarity of the P19 model is its strength at early times, which causes H photoionization by $z\simeq 14$, when minimal levels of
$\Omega_{\rm neutral} \sim $ few $10^{-4}$ and 
$\Omega_{\rm H_2} \sim 10^{-7}$ are reached.
Later on, with gas cooling re-balancing heating, some HI reforms and H$_2$ catches up.

\section{Impact of SNe}
\label{AppendixSN}

\begin{figure}
\includegraphics[width=0.4\textwidth]{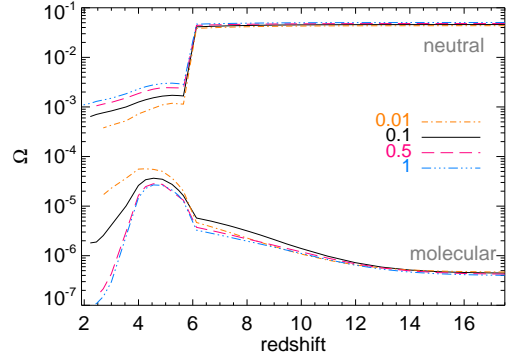}
\caption[]{\small
Redshift evolution of the neutral and molecular density parameters in models with SN efficiency of 
0.1 (black solid line, HM), 
0.01 (yellow dot-dashed line, HM-0.01), 
0.5 (magenta-red dashed line, HM-0.5) and 
1 (azure dot-dot-dot-dashed line, HM-1).
}
\label{fig:OmegaSN}
\end{figure}

\noindent 
In Fig.~\ref{fig:OmegaSN}, we show $\Omega_{\rm neutral}$ and $\Omega_{\rm H_2}$ as extracted from simulations including different SN efficiencies with a standard HM background, complementing our previous discussion of Fig.~\ref{fig:phaseFeedback}.
Efficiencies larger than the reference 0.1 value are suited for extremely powerful population-III sources, which would inhibit H$_2$ and star formation in the first few Gyr.
When compared to the reference 0.1 case, values of the molecular-gas density parameters, $\Omega_{\rm H_2}$, are lower (higher) for SN efficiencies of 0.5-1 (0.01).
That is a result of the more (less) efficient molecular-gas removal from star forming regions by stronger (weaker) feedback and the major (minor) destruction of H$_2$ into HI.
Correspondingly, $\Omega_{\rm neutral} $ values increase smoothly until $z\lesssim 6$ at slightly lower (higher) levels when weaker (stronger) feedback is active.
Qualitatively, these conclusions hold independently from the UV background (e.g. P19, which is not shown for clarity) and the variations found are comparable to those from other effects, such as cosmic-ray heating in star forming regions.

\section{The role of IMF and winds}
\label{AppendixWindsIMF}

\begin{figure}
\includegraphics[width=0.4\textwidth]{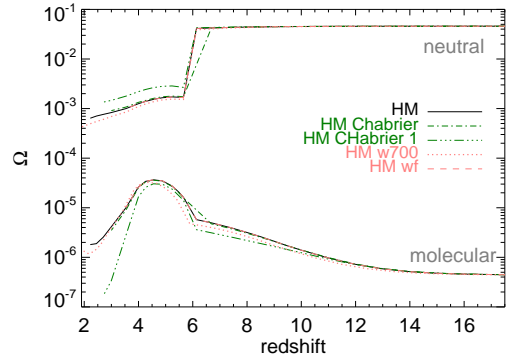}
\caption[]{\small
Redshift evolution of the neutral and molecular density parameters for the HM background set-up, having wind velocity of 350~km/s, wind delay time of 0.025 Hubble time and a Salpeter IMF (black solid lines, HM).
Results obtained by increasing,
the wind velocity to 700~km/s (pink dotted lines, HM-w700), 
the wind delay time to 0.1 Hubble time (pink dashed lines, HM-wf)
and by adopting a Chabrier IMF 
(green dot-dashed lines, HM-Chabrier) or 
a Chabrier IMF with unitary SN efficiency 
(green dot-dot-dot-dashed lines, HM-1-Chabrier)
are shown, too.
}
\label{fig:OmegaWindsIMF}
\end{figure}

\noindent
In Fig.~\ref{fig:OmegaWindsIMF}, we compare the effects of different IMFs for various feedback schemes.
We consider as reference a simulation with 
HM background,
Salpeter IMF, 
SN efficiency of 0.1,
350~km/s wind velocity and
wind delay time of 0.025 the Hubble time.
This simulation is re-run with variations including:
a Chabrier IMF, 
a Chabrier IMF with unitary SN efficiency,
a wind velocity of 700~km/s
and a longer wind delay time 0.1 the Hubble time.
Effects of a Chabrier IMF are small and results tend to follow the evolution of the corresponding runs with a Salpeter IMF.
The small effects due to changes of the IMF originate from the small difference in the injected entropy and the SN rates varying only by a factor of 1.4.
The model with Chabrier IMF and unitary SN efficiency features some departures from the reference for both $\Omega_{\rm neutral}$ and $\Omega_{\rm H_2}$, indicating a more relevant role of the feedback scheme.
In the case with wind velocity of 700~km/s, the results are very close to those of the reference model. 
Changing other wind parameters, such as the wind delay time to get re-coupled to hydrodynamic, has no dramatic effect.
Similar conclusions hold for different UV scenarios, too.

\section{Estimates of the dynamical time}
\label{AppendixDynamicalTime}

A test mass released from rest at distance $r$ from the center of the gravitational potential of a constant density field, $\rho$, under assumption of spherical symmetry, moves according to the equation of motion of a harmonic oscillator: 
\begin{equation}
\ddot r = - \frac{G M(r) }{r^2} = - \frac{ 4 \pi G \rho }{3} r,
\end{equation}
being $G$ the gravitational constant and $M(r)$ the mass enclosed within $r$.
The angular frequency $\omega $ satisfies the relation
$\omega^2 = 4\pi G \rho/3 $ and
the period is 
\begin{equation}
\label{Period}
T = \frac{ 2 \pi }{ \omega} = \sqrt{    \frac{ 3 \pi }{ G \rho } }.
\end{equation}
The test mass reaches the center in 1/4 the oscillation period, i.e. in a dynamical time, $t_{\rm dyn }$
\cite[e.g.][]{BinneyTremaine2008}: 
\begin{equation}
 t_{\rm dyn} 
 \equiv \frac{T}{4} 
 = \sqrt{ \frac{ 3 \pi }{ 16 G \rho} }.
\end{equation}
If, at any given redshift,
$ \rho \equiv \Delta \, \rho_{\rm crit}$,
the critical density is $\rho_{\rm crit} = 3 H^2 / (8\pi G) $, 
and the Hubble time is $t_{\rm H} = 1/H $, then
\begin{equation}
 t_{\rm dyn} 
 = \sqrt{ \frac{ 3 \pi }{ 16 G \Delta \rho_{\rm crit}} } 
 =  \frac{ \pi }{ \sqrt{2 \Delta} } ~ t_{\rm H},
\end{equation}
For $ \Delta \simeq 100$ up to 2000, $t_{\rm dyn}$ decreases from
$t_{\rm dyn} \simeq 0.2\,t_{\rm H}$ down to $ 0.05\,t_{\rm H}$.
A typical value of $ \Delta \simeq 500$ 
(corresponding to $ \sim 1800 $ times the mean matter density, or about   $ 10^4 $ times the mean baryon density) leads to $t_{\rm dyn} \simeq 0.1\,t_{\rm H}$.

\end{document}